\documentclass[12pt,a4paper]{article}

\usepackage{fullpage}

\usepackage{amssymb}
\usepackage{amsmath}
\usepackage{amsthm}
\usepackage{bm}
\usepackage[round,longnamesfirst]{natbib}
\usepackage{url}
\usepackage{graphicx}
\usepackage{multirow}
\usepackage{hyperref}
\usepackage{hypernat}
\usepackage{enumerate}
\usepackage{setspace}
\usepackage{booktabs,rotating}
\usepackage{multirow}

\title{A class of goodness-of-fit tests for spatial extremes models based on max-stable processes}

\author{
   Ivan Kojadinovic$^{1}$ \and Hongwei Shang$^{2}$ \and Jun Yan$^{2}$\\[5mm] 
   \small{$^{1}$Laboratoire de math\'ematiques et applications, UMR CNRS 5142} \\
   \small{Universit\'e de Pau et des Pays de l'Adour} \\
   \small{B.P. 1155, 64013 Pau Cedex, France} \\
   \small{\texttt{ivan.kojadinovic@univ-pau.fr}} \\[5mm] 
   \small{$^{2}$Department of Statistics} \\
   \small{University of Connecticut, 215 Glenbrook Rd. U-4120} \\
   \small{Storrs, CT 06269, USA} \\
   \small{\texttt{\{hongwei.shang,jun.yan\}@uconn.edu}}
 }

\date{}

\newcommand{\R}{\mathbb{R}}

\newcommand{\dd}{\mathrm{d}}
\newcommand{\Cb}{\mathbb{C}}
\newcommand{\CC}{\mathcal{C}}
\renewcommand{\AA}{\mathcal{A}}

\newcommand{\OO}{\mathcal{O}}
\newcommand{\XX}{\mathcal{X}}
\newcommand{\ZZ}{\mathcal{Z}}
\newcommand{\NN}{\mathcal{N}}

\newcommand{\Ex}{\mathrm{E}}
\newcommand{\Wb}{\mathbb{W}}

\renewcommand{\vec}{\bm}
\newcommand{\1}{\mathbf{1}}

\renewcommand{\th}{{\vec \theta}}
\newcommand{\vth}{{\vec \vartheta}}

\newcommand{\Ap}{\hat A_n^{\mathrm{P}}}
\newcommand{\Apc}{\hat A_{n,c}^{\mathrm{P}}}
\newcommand{\Acfg}{\hat A_n^{\mathrm{CFG}}}
\newcommand{\Acfgc}{\hat A_{n,c}^{\mathrm{CFG}}}
\newcommand{\Aht}{\hat A_n^{\mathrm{HT}}}

\newcommand{\Ep}{\hat \xi_{B,n}^{\mathrm{P}}}
\newcommand{\Ecfg}{\hat \xi_{B,n}^{\mathrm{CFG}}}
\newcommand{\Eht}{\hat \xi_{B,n}^{\mathrm{HT}}}

\newtheorem{prop}{Proposition}
\renewcommand{\qed}{$\blacksquare$}

\begin{document}
\maketitle

\begin{abstract}
Parametric max-stable processes are increasingly used to model spatial extremes. Starting from the fact that the dependence structure of a max-stable process is completely characterized by an extreme-value copula, a class of goodness-of-fit tests is proposed based on the comparison between a nonparametric and a parametric estimator of the corresponding unknown multivariate Pickands dependence function. Because of the high-dimensional setting under consideration, these functional estimators are only compared at a specific set of points at which they coincide, up to a multiplicative constant, with estimators of the extremal coefficients. The nonparametric estimators of the Pickands dependence function used in this work are those recently studied by Gudendorf and Segers. The parametric estimators rely on the use of the {\em pairwise pseudo-likelihood} which extends the concept of pairwise (composite) likelihood to a rank-based context. Approximate $p$-values for the resulting margin-free tests are obtained by means of a {\em one- or two-level parametric bootstrap}. Conditions for the asymptotic validity of these resampling procedures are given based on the work of Genest and R\'emillard. The finite-sample performance of the tests is investigated in dimension 10 under the Smith, Schlather and geometric Gaussian models. An application of the tests to rainfall data is finally presented.

\medskip


\medskip

\end{abstract}

\section{Introduction}
\label{intro}

The measurement of extremes, especially in the environment, is often spatial in nature as variables such as precipitation, temperature, pollutant concentration, or wind speed, are recorded over time at a network of sites. As rare events that occur at multiple locations simultaneously or within a very short time period can cause extensive damage, the modeling of spatial dependence in the analysis of extremes appears crucial from a risk management perspective. In contrast to univariate extreme-value theory which is rather mature and has been applied to a variety of fields \citep[see e.g.][for an overview]{Col01}, spatial extremes analysis gained sharpened focus only relatively recently \citep[see e.g.][]{Smi90, Sch02, SchTaw03, PadRibSis10, BlaDav11, DavGho12, DavPadRib12, FueHenRei13, Rib13}.

One natural approach for modeling spatial extremes consists of using {\em max-stable processes} as the latter stem from an extension of multivariate extreme-value theory to the process setting. Several parametric models were derived from so-called {\em spectral representations} of max-stable processes \citep[see e.g.][]{deH84,Sch02,deHPer06}. Among the most frequently encountered models, one finds the Smith, the Schlather and the class of Brown-Resnick models. The recent literature suggests to base the inference about the parameters of these models on the {\em pairwise likelihood} because the full likelihood is typically intractable \citep[see e.g.][for more details on this matter]{PadRibSis10, GenMaSan11, DavGho12}.

The quality of the fit of a spatial model based on a parametric max-stable process seems to have been essentially investigated by means of graphical tools. \citet{Smi90} proposed to compare nonparametric with parametric estimates of pairwise and higher-order {\em extremal coefficients}. The latter coefficients describe the spatial dependence among the sites as explained for instance in \citet{SchTaw03}. When restricted to pairs of sites, the approach proposed by \cite{Smi90} consists of standardizing, for each pair of sites, the difference between a nonparametric and a parametric estimate of the corresponding pairwise extremal coefficient by the jackknife standard error of the nonparametric estimate. %In the case of pairwise extremal coefficients, the parametric estimates can be directly computed from on the estimated model parameters since the bivariate marginal density is known in close-form. 
The standardized differences for all pairs of sites can be plotted against the corresponding parametric estimates of the pairwise extremal coefficients or against the distances between the sites. This provides a visual check similar to a residual plot for linear regression. %For higher-order extremal coefficients, the parametric estimates have to be approximated from simulations. 
An alternative graphical approach was used in \citet{PadRibSis10} and \citet{DavGho12} who assessed the fit of models for various subsets of sites with a particular quantile-quantile plot. Specifically, for a given subset of sites, the annual maximum over the subset was obtained for each of the observed years, forming the sample quantiles of the annual maxima for that subset. These sample quantiles were then plotted against population quantiles approximated from a large number of datasets generated from the fitted model. The described approach is a multivariate extension of the graphical diagnostic tool used in the analysis of univariate extremes \citep[see e.g.][]{Col01}. 

In the case of a clear departure from the hypothesized model, such graphical approaches are known to be useful tools that can help a user better understand the underlying characteristics of the data. Deciding to reject (or not) a model on the basis of the available graphs can however turn out to be a very subjective process as, among other things, the perceived departure from the hypothesized model depends on the sample size. For that reason, it is frequently advised to use such graphical tools in conjunction with formal testing procedures \citep[see e.g.\ the discussion in][]{Dag86}.   

To the best of our knowledge, no formal goodness-of-fit tests have been developed for spatial models based on max-stable processes. The purpose of this work is to fill this gap. Starting from the well-known fact that the dependence structure of a max-stable process is uniquely characterized by an {\em extreme-value copula} \citep[see e.g.][]{GudSeg10,DavPadRib12,RibSed13}, it would seem natural to base goodness-of-fit tests for the spatial models under consideration on goodness-of-fit tests for copulas. The latter tests received a lot of attention in the recent literature \citep[see e.g.][]{GenRem08,GenRemBea09,KojYanHol11}. They were adapted to deal specifically with bivariate extreme-value copulas by \cite{GenKojNesYan11} who derived test statistics from the empirical process comparing a nonparametric estimator with a parametric estimator of the so-called {\em Pickands dependence function} uniquely defining the underlying extreme-value copula \citep[see e.g.][]{GudSeg10}. A straightforward extension of these bivariate tests to the current large-dimensional setting does not however appear computationally feasible. 
 
The tests proposed in this work exploit the idea initially proposed by \cite{Smi90} consisting of comparing nonparametric estimators of extremal coefficients with parametric estimators to assess the fit of a model. Because extremal coefficients can be expressed in terms of the Pickands dependence function, the derived tests can also be cast in the framework considered by \cite{GenKojNesYan11}. More specifically, the tests are based on the absolute or squared differences between nonparametric and parametric rank-based estimators of extremal coefficients. The rank-based nature of the estimators implies that the tests are margin-free, which is a desirable feature. The nonparametric estimators are the two multivariate rank-based estimators of the Pickands dependence function recently studied by \cite{GudSeg12}. The parametric estimators rely on the estimation of the parameters of the hypothesized model using the {\em pairwise pseudo-likelihood} which extends the concept of pairwise (composite) likelihood considered for instance in \citet{PadRibSis10} and \citet{DavGho12} to the current rank-based setting. If closed form expressions for the parametric extremal coefficients exist, approximate $p$-values for the tests can be obtained using a classical (one-level) parametric bootstrap. If such expressions do not exist for the hypothesized model, the parametric estimators are obtained by applying the nonparametric estimators to a large sample generated from the fitted model and a {\em two-level} parametric bootstrap is used to compute approximate $p$-values. In both cases, the asymptotic distribution of the test statistics under the null is obtained and technical conditions under which the previous testing procedures are valid in the sense of Theorems~1 and~2 of \cite{GenRem08} are derived. Although the finite-sample performance of the tests is only investigated in the case of three frequently encountered models, the derived procedures can in principle be used to assess the fit of most other parametric max-stable processes. 

The paper is organized as follows. Section~\ref{spatial} contains a brief and partial overview of spatial models based on max-stable processes and recalls the connections existing between extremal coefficients and copulas. The third section is devoted to a detailed presentation of the proposed testing procedures. Section~\ref{simulations} partially reports the results of a Monte Carlo experiment involving 10 sites and 50, 100 or 200 observations per site. The last section presents the application of the tests to the Swiss rainfall data analyzed in \citet{DavPadRib12}.% and contains concluding remarks. 

%Note finally that the code of all the tests studied in this work will be documented and released as an R package accompanying the paper.

\section{Spatial models based on max-stable processes}
\label{spatial}

%Max-stable processes \citep[see e.g.][]{deH84,deHPer06} arise as natural tools for modeling spatial maxima \citep[see e.g.][]{SchTaw03,PadRibSis10,DavGho12,DavPadRib12}. 

\subsection{Max-stable processes}
\label{maxstable}

Let $\XX$ be a contiguous subset of $\R^2$ containing $\vec o = (0,0)$ and representing a spatial domain of interest. For convenience, we shall focus on stochastic processes on $\XX$ that have unit Fr\'echet margins. A process from this class will be denoted by $Z$ and regarded as a random function $Z:\XX \to \R$ as we continue. We therefore have that, by definition, for any site $\vec x \in \XX$,
$$
\Pr\{ Z(\vec x) \leq z \} = \exp(-1/z), \qquad z > 0.
$$
A process $Z$ on $\XX$ with unit Fr\'echet margins is {\em max-stable} if it satisfies the max-stability property, i.e., for any $\{\vec x_1,\dots,\vec x_d\} \subset \XX$ and any integer $k > 0$,
$$
\Pr\{ Z(\vec x_1) \leq z_1,\dots,Z(\vec x_d) \leq z_d \} = \Pr\{ Z(\vec x_1) \leq k z_1,\dots,Z(\vec x_d) \leq k z_d \}^k, \quad z_1,\dots,z_d > 0.
$$
The max-stability property implies that the higher-order marginal distributions of $Z$ belong to the class of multivariate extreme-value distributions; see for instance \cite{DavPadRib12}, \cite{Rib13} and the references therein for a more detailed introduction.

Families of parametric max-stable processes used in applications were derived from so-called {\em spectral representations}. A first such representation is due to \cite{deH84} \citep[see also e.g.][Section 2]{Rib13} and states that, for any stationary max-stable process $Z$ on $\XX$ with continuous sample paths and unit Fr\'echet margins, there exists a non-negative continuous function $f$ on $\R^4$ satisfying two conditions to be given below, such that $Z$ has the same distribution as the process on $\XX$ defined by
\begin{equation}
\label{specrep1}
\vec x \mapsto \sup_{j \geq 1} S_j f(\vec x,\vec L_j),
\end{equation}
where $(S_1,\vec L_1), (S_2,\vec L_2), \dots$ are the points of a Poisson process on $(0,\infty) \times \R^2$ with intensity $\dd s/s^2 \times \dd \vec \ell$. The function $f$ in~\eqref{specrep1} satisfies $\int_{\R^2} f(\vec x, \vec y) \dd \vec y = 1$ for all $\vec x \in \R^2$ and $\int_{\R^2} \sup_{\vec x \in K} f(\vec x, \vec y) \dd \vec y < \infty$ for all compact sets $K \subset \XX$. 

A class of rainfall storm models is obtained by defining $f$ in~\eqref{specrep1} as $f(\vec x,\vec y) = g(\vec x - \vec y)$, where $g$ is a bivariate probability density function (p.d.f.) on $\R^2$. In this model, $S_j g(\vec x - \vec L_j)$ can be interpreted as the impact at location $\vec x$ of a storm of intensity $S_j$ centered at location $\vec L_j$, and $\sup_{j \geq 1} S_j g(\vec x - \vec L_j)$ as the impact of the strongest such episode experienced at $\vec x$. The case when $g$ is taken equal to $\phi^{(2)}_{\bm \Sigma}$, the bivariate normal p.d.f.\ with mean zero and covariance matrix ${\bm \Sigma}$, was considered by \cite{Smi90} and is therefore frequently referred to as the {\em Smith model} \citep[see also e.g.][]{Col93,deHPer06,PadRibSis10}. The process in~\eqref{specrep1} being stationary, its bivariate marginal distributions are fully described by the cumulative distribution function (c.d.f.) of $(Z(\vec o),Z(\vec x))$, where $\vec o \in \XX$ is the origin and $\vec x$ is an arbitrary site in $\XX$. From \citet{Smi90} \citep[see also][Appendix A.3]{PadRibSis10}, we have that, for any $z_1,z_2>0$,
\begin{equation}
\label{smith_bivariate}
\Pr \{ Z(\vec o)\leq z_{1}, Z(\vec x)\leq z_{2} \}
= \exp \left\{-\frac{1}{z_1}\Phi\left(\frac{a}{2}+\frac{1}{a} \log \frac{z_{2}}{z_{1}}\right)-\frac{1}{z_2}\Phi\left(\frac{a}{2}+\frac{1}{a}\log\frac{z_{1}}{z_{2}}\right)\right\},
\end{equation} 
where $\Phi$ is the univariate standard normal c.d.f.\ and $a^2= \vec x^\top \ {\bm \Sigma}^{-1} \vec x$.

A second key spectral representation is due to \cite{Sch02}. Following \cite{DavPadRib12} and \citet[Section 2]{Rib13}, let $S_1,S_2,\dots$ be the points of a Poisson process on $(0,\infty)$ of intensity $\dd s/s^2$. Then, for any stationary max-stable process $Z$ on $\XX$ with continuous sample paths and unit Fr\'echet margins, there exists a positive stochastic process $W$ on $\R^2$ with continuous sample paths and $\Ex\{W(\vec x)\}=1$ for all $\vec x \in \R^2$ such that $Z$ has the same distribution as the process on $\XX$ defined by
\begin{equation}
\label{specrep2}
\vec x \mapsto \sup_{j \geq 1} S_j W_j(\vec x),
\end{equation}
where $W_1,W_2,\dots$ are independent copies of $W$. 

Starting from~\eqref{specrep2}, another spatial model frequently encountered in the literature was proposed by \citet{Sch02} and consists of defining $W_j$ as $W_j(\vec x) = \max\{0, \sqrt{2\pi} \epsilon_j(\vec x)\}$, where $\epsilon_1,\epsilon_2,\dots$ are independent copies of a stationary Gaussian process $\epsilon$ on $\R^2$ with unit variance and correlation function $\rho$. For this model, frequently referred to as the {\em Schlather model}, we have, for any $z_1,z_2>0$, that
\begin{equation}
\label{schlather_bivariate}
\Pr \{ Z(\vec o)\leq z_{1}, Z(\vec x)\leq z_{2} \}
= \exp \left\{-\frac{1}{2} \left( \frac{1}{z_1} + \frac{1}{z_2} \right) \left( 1 + \left[1 - \frac{2 \{ \rho(\vec x) + 1\}z_1 z_2 }{(z_1 + z_2)^2} \right]^{1/2} \right)\right\}.
\end{equation} 
As is well-known, a drawback of this model is that it cannot model spatial independence between sites. Extensions of the Schlather model are discussed for instance in \citet{DavGho12} and in \cite{Rib13}.

A third spatial model that shall be considered in this work is the so-called {\em geometric Gaussian process}. It is obtained by defining $W_j$ in~\eqref{specrep2} as 
\begin{equation}
\label{geogaussian}
W_j(\vec x) = \exp\{ \sigma \epsilon_j(\vec x) - \sigma^2/2\},
\end{equation}
where $\sigma > 0$ and $\epsilon_1,\epsilon_2,\dots$ are independent copies of a stationary Gaussian process $\epsilon$ on $\R^2$ with unit variance and correlation function $\rho$. For this model, the c.d.f.\ of $(Z(\vec o), Z(\vec x))$ is given by~\eqref{smith_bivariate} but with $a^2 = 2 \sigma^2 \{1 - \rho(\vec x)\}$. Note that this process is a particular Brown-Resnick process \citep{DavRes84,KabSchdeH09}. We did not consider the latter class of models in our Monte Carlo experiment as random number generation from these processes can be tricky as discussed for instance in \citet[Section 7]{Rib13} \citep[see also][]{OesKabSch12}.

\subsection{Extremal coefficients}

As explained for instance in \citet{SchTaw03} or \citet{DavGho12}, a natural way of measuring dependence among spatial maxima modeled by a max-stable process $Z$ on $\XX$ with unit Fr\'echet margins consists of examining the distribution of $\sup_{\vec x \in \XX} Z(\vec x)$, which can be expressed as 
$$
\Pr \{ \sup_{\vec x \in \XX} Z(\vec x) \leq z \} %= \Pr \{ Z(\vec x) \leq z, \mbox{ for all } \vec x \in \XX \} 
= \exp(- \xi_\XX / z ), \qquad z > 0,
$$
in terms of the {\em extremal coefficient}  $\xi_\XX$ of the set $\XX$ \citep[see e.g.][for more details]{DavGho12}. If $\xi_\XX$ is close to one, then the distribution of $\sup_{\vec x \in \XX} Z(\vec x)$ is close, for any $\vec x \in \XX$, to that of the random variable $Z(\vec x)$ (which is unit Fr\'echet by definition), thereby indicating almost perfect dependence between the spatial maxima. Weaker dependence between the maxima yields larger values of $\xi_\XX$.

Similarly, with the notation $D = \{1,\dots,d\}$, the extremal coefficient of a set of locations $\{\vec x_i : i \in D\} \subset \XX$, $d \geq 2$, is defined through the following equation:
\begin{equation}
\label{theta_D}
\Pr \left\{ \max_{j \in D} Z(\vec x_j) \leq z \right\} = \Pr \{ Z(\vec x_1) \leq z, \dots, Z(\vec x_d) \leq z \} = \exp(- \xi_D / z ), \qquad z > 0.
\end{equation}
It is easy to verify that independence among the maxima measured at $\vec x_1,\dots,\vec x_d$ yields $\xi_D = d$, while perfect dependence gives $\xi_D = 1$. More generally, $\xi_D \in [1,d]$.

The extremal coefficient $\xi_D$ can be expressed in terms of the so-called {\em Pickands dependence function} of the random vector $(Z(\vec x_1), \dots, Z(\vec x_d))$. Indeed, the random vector $(Z(\vec x_1), \dots, Z(\vec x_d))$ having continuous margins, its c.d.f.\ can be uniquely expressed \citep{Skl59} as
\begin{equation}
\label{sklar}
C\{F(z_1),\dots,F(z_d)\}, \qquad z_1,\dots,z_d > 0,
\end{equation}
where $F$ is the c.d.f.\ of the unit Fr\'echet distribution and $C$ is a {\em copula} that is of the {\em extreme-value} type \cite[see e.g.][]{GudSeg10}. The copula $C$ is simply the c.d.f.\ of the random vector $(F\{Z(\vec x_1)\}, \dots, F\{Z(\vec x_d)\})$. Because it is of the extreme-value type, $C$ can in turn be expressed as
\begin{equation}
\label{extreme_value_copula}
C(\vec u) = \exp \left\{ \left( \sum_{j=1}^d \log u_j \right) A \left(\frac{\log u_1}{\sum_{j=1}^d \log u_j}, \dots, \frac{\log u_d}{\sum_{j=1}^d \log u_j} \right) \right\}, \,\vec u \in (0,1]^d \setminus \{(1,\dots,1)\},
\end{equation}
where $A:\Delta_{d-1} \to [1/d,1]$ is the {\em Pickands dependence function} and $\Delta_{d-1} = \{(w_1,\dots,w_d) \in [0,1]^{d-1} : w_1 + \dots + w_d = 1 \}$ is the unit simplex \citep[see e.g.][for more details]{GudSeg12}.

Combining expression~\eqref{extreme_value_copula} with~\eqref{sklar} and equating it to~\eqref{theta_D}, one obtains that $\xi_D = d A(1/d,\dots,1/d)$. More generally, it can be verified that the extremal coefficient of any subset of sites $\{\vec x_i : i \in B\}$ with $B \subset D$, $|B| \geq 2$, can be expressed as
\begin{equation}
\label{extremal_Pickands}
\xi_B = |B| A(\vec w_B),
\end{equation}
where $\vec w_B$ is the vector of $\Delta_{d-1}$ such that $w_{B,i} = 1/|B|$ if $i \in B$ and $w_{B,i} = 0$ otherwise. Thus, the set of extremal coefficients $\xi_B$, $B \subset D$, $|B| \geq 2$, merely corresponds to the scaled values of the Pickands dependence function $A$ at the points $\vec w_B$, $B \subset D$, $|B| \geq 2$, of $\Delta_{d-1}$. As is well-known, it therefore clearly appears that the set of extremal coefficients $\xi_B$, $B \subset D$, $|B| \geq 2$, does not fully characterize the extreme-value copula $C$. Properties of the set of extremal coefficients are studied in \citet{SchTaw03}.

\section{Goodness-of-fit tests based on extremal coefficients}
\label{description}

Let the random variables $X_1,\dots,X_d$ represent the maxima of a quantity of interest (such as temperature or precipitation) at the $d$ locations in $\{\vec x_1,\dots,\vec x_d\} = \{\vec x_i : i \in D\} \subset \XX$ over a period $T$ (typically a year), and assume that the unknown c.d.f.\ of the random vector $(X_1,\dots,X_d)$ belongs to the class of multivariate extreme-value distributions. It follows that, for any $j \in D$, the unknown univariate c.d.f.\ $F_j$ of $X_j$ belongs to the class of generalized extreme-value distributions and that $Z_j = - 1 / \log \{ F_j(X_j) \}$ has a unit Fr\'echet distribution.

Consider further a parametric class $\ZZ = \{Z_\th : \th \in \OO \}$ of max-stable processes on $\XX$ with unit Fr\'echet margins, where $\OO$ is an open subset of $\R^p$ for some integer $p > 0$. We then know from the previous section that there exists a parametric family of copulas $\CC = \{C_\th : \th \in \OO \}$ such that, for any $\th \in \OO$, $C_\th$ is the copula of the random vector $(Z_\th(\vec x_1),\dots,Z_\th(\vec x_d))$. Because $\CC$ is a family of extreme-value copulas, $\CC$ can be defined from a parametric family of Pickands dependence functions $\AA = \{A_\th : \th \in \OO \}$ through~\eqref{extreme_value_copula}.

Let $C$ be the unknown extreme-value copula of $(X_1,\dots,X_d)$ and let $A$ be the corresponding unknown Pickands dependence function. Having at hand $n$ independent copies $(X_{1,1},\dots,X_{1,d}), \dots, (X_{n,1},\dots,X_{n,d})$ of the random vector $(X_1,\dots,X_d)$, we wish to test 
\begin{equation}
\label{H0}
H_0:C \in \CC \, (\mbox{i.e., } A \in \AA) \qquad \mbox{against} \qquad H_1:C \not \in \CC \, (\mbox{i.e., } A \not \in \AA).
\end{equation}
The rejection of $H_0$ will be interpreted as evidence in the data that the family of max-stable processes $\ZZ$ does not constitute an appropriate model.
 
As discussed in \cite{GenKojNesYan11}, a seemingly natural approach to the goodness-of-fit problem stated in~\eqref{H0} consists of comparing a nonparametric estimator $\hat A_n$ of the Pickands dependence function $A$ with a parametric estimator of $A$ under the null, both computed from $(X_{1,1},\dots,X_{1,d}), \dots, (X_{n,1},\dots,X_{n,d})$. The null hypothesis implies that there exists an unknown $\th_0 \in \OO$ such that $A = A_{\th_0}$. Given an estimator $\hat \th_n$ of $\th_0$, a natural estimator of $A$ under the null is therefore simply $A_{\hat \th_n}$. Such an approach was adopted in a bivariate context by \citet{GenKojNesYan11} who considered Cram\'er--von Mises test statistics derived from empirical processes on $\Delta_1$ of the form $\sqrt{n} (\hat A_n - A_{\hat \th_n})$.  

Because spatial problems usually involve a large number of sites $d$, a direct extension of the previous approach does not appear practically feasible as it would involve numerical integration over $\Delta_{d-1}$. Instead of comparing $\hat A_n$ with $A_{\hat \th_n}$ over the whole of $\Delta_{d-1}$, one possibility, as suggested by~\eqref{extremal_Pickands}, consists of considering a finite number of points in $\Delta_{d-1}$ such as the points $\vec w_B$, $B \subset D$, $|B| \geq 2$ and in defining
\begin{equation}
\label{generic}
S_{B,n} = \sqrt{n} |B| \left| \hat A_n(\vec w_B) - A_{\hat \th_n}(\vec w_B) \right|, \qquad B \subset D, |B| \geq 2.
\end{equation}
For a subset $B \subset D$ with $|B| \geq 2$, $S_{B,n}$ is nothing else than the scaled absolute difference between a nonparametric estimator of the extremal coefficient $\xi_B$ and a parametric estimator of the latter under the null. If the null hypothesis defined in~\eqref{H0} holds, then, clearly, so does the hypothesis
$$
H_{0,B}: \xi_B \in \{\xi_{B,\th} = |B| A_\th(\vec w_B) : \th \in \OO \}.
$$
The converse is however false in general. It follows that tests based on $S_{B,n}$ will not be consistent with respect to the hypotheses given in~\eqref{H0}.

In our simulations whose results will be partially reported in Section~\ref{simulations}, we focused on test statistics of the following form: $\sum_{B \subset D, |B| = 2} \{S_{B,n}\}^\alpha$, $\sum_{B \subset D, |B| = 3} \{S_{B,n}\}^\alpha$, $S_{D,n}$, $\sum_{B \subset D, |B| = \{2,3,d\}} \{S_{B,n}\}^\alpha$ and $\sum_{B \subset D, |B| = \{2,3,d\}} \{S_{B,n} / |B|\}^\alpha$, for $\alpha \in \{1,2\}$. The first four are based on the comparison of a nonparametric and a parametric estimator of extremal coefficients of various subsets of sites, while the last, through the division by $|B|$, compares the corresponding estimators of the unknown Pickands dependence function. %For $\alpha=2$, the statistics are Cram\'er--von Mises-like statistics.

\subsection{Nonparametric estimators of the Pickands dependence function}
\label{nonparametric}

In the realistic situation where the margins $F_1,\dots,F_d$ of $(X_1,\dots,X_d)$ are unknown, two nonparametric estimators of the unknown Pickands dependence function $A$ were recently derived by \citet{GudSeg12} as extensions of those proposed by \cite{GenSeg09} in the bivariate case. They are the rank-based versions of two well-known estimators of $A$, namely the Pickands estimator \citep{Pic81} and the Cap\'era\`a--Foug\`eres--Genest estimator \citep{CapFouGen97}. The latter will be abbreviated as CFG in the sequel. 

Let $\hat {\vec U}_i = (\hat U_{i,1},\dots, \hat U_{i,d})$, $i \in \{1,\dots,n\}$, be pseudo-observations computed from the available data by $\hat U_{i,j} = R_{i,j}/(n+1)$, where $R_{i,j}$ is the rank of $X_{i,j}$ among $X_{1,j},\dots,X_{n,j}$. The pseudo-observations can equivalently be rewritten as $\hat U_{i,j} = n \hat F_j(X_{i,j})/(n+1)$, where $\hat F_j$ is the empirical c.d.f.\ computed from $X_{1,j},\dots,X_{n,j}$, and where the scaling factor $n/(n+1)$ is classically introduced to avoid problems at the boundary of~$[0,1]^d$.

Let
$$
\hat \zeta_i(\vec w) = \bigwedge_{j=1}^d \frac{- \log \hat U_{i,j}}{w_j}, \qquad \vec w \in \Delta_{d-1}, \qquad i \in \{1,\dots,n\},
$$
where $\wedge$ denotes the minimum. The rank-based version of the Pickands and CFG estimators are then respectively defined by
$$
\Ap(\vec w) = 1 \Big/ \frac{1}{n} \sum_{i=1}^n \hat \zeta_i(\vec w), \quad \mbox{and} \quad \Acfg(\vec w) = \exp \left[ - \beta - \frac{1}{n} \sum_{i=1}^n \log \hat \zeta_i(\vec w) \right], \quad \vec w \in \Delta_{d-1},
$$
where $\beta = - \int_0^\infty \log(x) e^{-x} \dd x \approx 0.577$ is the Euler--Mascheroni constant.

From the above definitions, it is easy to verify that $\Ap(\vec e_1) = \dots = \Ap(\vec e_d)$ and that $\Acfg(\vec e_1) = \dots = \Acfg(\vec e_d)$, where $\vec e_1,\dots, \vec e_d$ are the standard basis vectors of $\R^d$. To ensure that the endpoint constraints $\Ap(\vec e_j) = \Acfg(\vec e_j) = 1$, $j \in D$, are satisfied, the previous estimators can be corrected as
$$
1/\Apc (\vec w) = 1/\Ap (\vec w) - 1/\Ap (\vec e_1) + 1, \qquad \vec w \in \Delta_{d-1},
$$
and
$$
\log \Acfgc (\vec w)  = \log \Acfg (\vec w) - \log \Acfg (\vec e_1), \qquad \vec w \in \Delta_{d-1},
$$
respectively. These corrections were suggested in \citet{GudSeg12} as natural extensions of those proposed in the case of known margins for $d=2$ by \citet{Deh91} and \citet{CapFouGen97}, respectively.

In the bivariate case, the above corrected versions were found to behave better than the uncorrected versions in small samples in \citet{GenSeg09} and \citet{GenKojNesYan11}. As verified in \citet{GudSeg12}, $\Ap$ and $\Apc$ (resp.\ $\Acfg$ and $\Acfgc$) become indistinguishable as $n$ tends to infinity. Also, $\Acfgc$ was found, overall, to outperform $\Apc$ in several bivariate Monte-Carlo experiments \citep[see e.g.][]{GenSeg09,KojYan10c}. The same empirical conclusion was obtained by \cite{GudSeg12} in dimension three.

A second corrected version of the Pickands estimator is obtained when considering, in the current rank-based context, the estimator initially proposed by \citet{HalTaj00} for $d=2$ and known margins . It is given by
$$
\Aht(\vec w) = \Ap(\vec w) / \Ap(\vec e_1),\qquad \vec w \in \Delta_{d-1}.
$$
  
By analogy with~\eqref{extremal_Pickands}, for any $B \subset D$, $|B| \geq 2$, these three corrected estimators give three estimators of the extremal coefficient $\xi_B$ as
\begin{equation}
\label{extremal_estimators}
\Ep = |B| \Apc(\vec w_B), \quad\Eht = |B| \Aht(\vec w_B), \quad \mbox{and} \quad \Ecfg = |B| \Acfgc(\vec w_B).
\end{equation}

Note that a multivariate rank-based version of the estimator suggested by \citet{Smi90} can be expressed as $|B| \Ap(\vec w_B)$ with our notation. The estimator $\Ep = |B| \Apc(\vec w_B)$ considered in this work is therefore merely a corrected version of the latter. Furthermore, the estimator $\Eht$ is nothing else than the so-called {\em naive estimator} proposed by \citet{SchTaw03} with threshold $z=0$ when computed from the transformed pseudo-observations $- 1/\log \hat U_{i,j}$, $i \in \{1,\dots,n\}$, $j \in B$. To see this, it suffices to start from the log likelihood given in \citet[Section 4.2]{SchTaw03}, set its derivative to zero and solve for the extremal coefficient.

The fact that the estimators of $\xi_B$ given in~\eqref{extremal_estimators} are defined from corrected estimators of the Pickands dependence function does not ensure that they are restricted to the range $[1,|B|]$. Hence, as suggested in \cite{SchTaw03}, it might be necessary to truncate them to the range $[1,|B|]$ in the case of small samples. In our experiments however, we have not observed the need for such a truncation.  

From a practical perspective, note finally that the estimators $\Apc$, $\Aht$ and $\Acfgc$ were implemented and are available in the \texttt{copula} package \citep{copula} of the \textsf{R} statistical environment \citep{Rsystem}

\subsection{Estimators of the Pickands dependence function under the null}

Recall that $C$ denotes the unknown copula of $(X_1,\dots,X_d)$ and that the null hypothesis states that there exists $\th_0 \in \OO$ such that $C=C_{\th_0}$. As proposed by \citet{GenGhoRiv95}, a natural way of estimating $\th_0$ under the null in the rank-based context under consideration would be to maximize the log {\em pseudo-likelihood} 
$$
\ell(\th) = \sum_{i=1}^n \log c_\th \left( \hat U_{i,1},\dots,\hat U_{i,d} \right), \qquad \th \in \OO,
$$
where $c_\th$ is the p.d.f.\ associated with $C_\th$ and where the term {\em pseudo} in {\em pseudo-likelihood} refers to the fact that the p.d.f.\ is evaluated at the pseudo-observations $\hat{\vec U}_1,\dots,\hat{\vec U}_n$. However, because of the extreme-value nature of the copula $C_\th$, a combinatorial explosion occurs when one attempts to compute $c_\th = \partial^d C_\th / \partial u_1 \dots \partial u_d$ \citep[see e.g.\ the discussion in][end of Section 2.1]{DavGho12}. It follows that, for most parametric max-stable processes of practical interest in spatial statistics, the maximization of the log pseudo-likelihood is practically unfeasible for $d > 4$ \citep[see][for recent results concerning the Smith model]{GenMaSan11}. As discussed for instance in \citet{DavGho12} or \citet{PadRibSis10}, an alternative consists of using a {\em composite likelihood} approach \citep{Lin88}, which, in the pseudo-likelihood context under consideration, yields the {\em pairwise log pseudo-likelihood}
\begin{equation}
\label{composite}
\tilde \ell(\th) = \sum_{i=1}^n \sum_{\{j,k\} \subset D} \log c_\th^{(j,k)} \left( \hat U_{i,j},\hat U_{i,k} \right),\qquad \th \in \OO,
\end{equation}
where $c_\th^{(j,k)}$ is the p.d.f.\ of the copula of the bivariate random vector $(Z_\th(\vec x_j),Z_\th(\vec x_k))$ for a pair of sites $\{\vec x_j,\vec x_k\}$. Note that the efficiency of the maximum pairwise pseudo-likelihood estimator might be increased by restricting the expression above to pairs of sites that are closer than a specified threshold as empirically illustrated in \citet{PadRibSis10} in the case of the maximum pairwise likelihood estimator. 

For any $\th \in \OO$, recall that $A_\th$ denotes the Pickands dependence function associated with the extreme-value copula $C_\th$. Furthermore, let $\hat \th_n$ be the maximizer of~\eqref{composite}. An estimator of the Pickands dependence function under the null $A_{\th_0}$ is then given by $A_{\hat \th_n}$. For a given $B \subset D$, $|B| \geq 2$, it follows that an estimator of the extremal coefficient $\xi_B$ under the null is given by 
\begin{equation}
\label{estimator_theta_null}
\xi_{B,\hat \th_n} = |B| A_{\hat \th_n}(\vec w_B).
\end{equation}

The previous estimator can however only be computed if a closed form expression for $A_\th$ is available. If it is not the case, $\xi_B$ can be estimated provided one knows how to generate a random sample from $C_\th$. For some fixed real $\gamma > 0$ (typically much greater than one), let $m = \lfloor \gamma n \rfloor$ and let $(V_{1,1},\dots,V_{1,d}),\dots,(V_{m,1},\dots,V_{m,d})$ be a random sample of size $m$ from $C_{\hat \th_n}$, independent of the available data conditionally on $\hat \th_n$. Then, an estimator of $\xi_B$ is
\begin{equation}
\label{estimator_theta_null_2}
\hat \xi_{B,\hat \th_n,m} = |B| \hat A_m(\vec w_B), 
\end{equation}
where $\hat A_m$ is one of the three corrected nonparametric estimators of the Pickands dependence function defined in Section~\ref{nonparametric} computed from the pseudo-observations obtained from $(V_{1,1},\dots,V_{1,d}),\dots,(V_{m,1},\dots,V_{m,d})$. 

To illustrate the use of the two estimators of $\xi_B$ under the null discussed above, we first consider the situation when $\ZZ = \{Z_\th : \th \in \OO \}$ corresponds to the Smith model, and then, for instance, to the Schlather model. Clearly, the estimator given in~\eqref{estimator_theta_null_2} has the highest applicability across models as it mostly relies on the availability of random number generation routines.

\subsubsection{The case of the Smith model}
\label{Smith}

For the Smith model, it is known \citep[see e.g.][page 147]{SchTaw03} that $C_\th$ is a $d$-dimensional H\"usler--Reiss copula \citep{HusRei89}. The dependence in this model is controlled by the covariance matrix $\bm \Sigma$ of the bivariate standard normal p.d.f.\ $\phi^{(2)}_{\bm \Sigma}$ which determines the elliptical contour of a typical storm. In other words, $\th = \bm \Sigma$. 

Starting from~\eqref{smith_bivariate}, one recovers that the Pickands dependence function of the random vector $(Z_\th(\vec x_j),Z_\th(\vec x_k))$, for a pair of sites $\{\vec x_j, \vec x_k\}$, is given by  
\begin{equation}
\label{smith_bivariate_pickands}
A_\th(\vec w) = w_1 \Phi\left(\frac{a_{\{j,k\}}}{2}+\frac{1}{a_{\{j,k\}}} \log \frac{w_1}{w_2}\right) + w_2 \Phi\left(\frac{a_{\{j,k\}}}{2}+\frac{1}{a_{\{j,k\}}}\log\frac{w_2}{w_1}\right), \qquad \vec w \in \Delta_1,
\end{equation}
where $\Phi$ is the standard normal c.d.f.\ and $a_{\{j,k\}}^2= (\vec x_j - \vec x_k)^\top \ {\bm \Sigma}^{-1}(\vec x_j - \vec x_k)$. The previous expression can be used to obtain the expression of the bivariate H\"usler--Reiss copula through~\eqref{extreme_value_copula}, and the expression of the bivariate p.d.f.\ $c^{(j,k)}_\th$ needed in~\eqref{composite}. As one can see, the dependence is controlled by the distance $a_{\{j,k\}}$ between sites $\vec x_j$ and $\vec x_k$, which depends on $\th = {\bm \Sigma}$.

The c.d.f.\ of the $d$-dimensional H\"usler--Reiss copula was recently expressed in a convenient form by \citet{NikJoeLi09} using as parameter a symmetric $d \times d$ matrix with off-diagonal elements $\delta_{ik} > 0$ and $\delta_{ii}^{-1} = 0$, such that any $(d-1) \times (d-1)$ matrix ${\bm \Gamma}_j$ with element $(i,k)$ given by
$$
\frac{\delta_{ij}^{-2} + \delta_{kj}^{-2} - \delta_{ik}^{-2}}{ 2\delta_{ij}^{-1}\delta_{kj}^{-1}}, \qquad i,k \in D \setminus \{j\},
$$
is a correlation matrix. Let $\Phi^{(d-1)}_{{\bm \Gamma}_j}$ be the c.d.f.\ of the $(d-1)$-dimensional standard normal distribution with correlation matrix ${\bm \Gamma}_j$. From the work of \citet{NikJoeLi09}, we have that the Pickands dependence function of the $d$-dimensional H\"usler--Reiss copula parametrized by the matrix $(\delta_{ij})_{i,j \in D}$ is 
$$
A_\delta(\vec w) = \sum_{j=1}^d {w_j \Phi^{(d-1)}_{{\bm \Gamma}_j} \left( \delta_{ij}^{-1} + \frac{\delta_{ij}}{2} \log \frac{w_j}{w_i} : i \in D \setminus \{j\} \right)}, \qquad \vec w \in \Delta_{d-1}.
$$
Setting all but two $w_j$ to zero to obtain a bivariate Pickands dependence function and comparing the resulting function with~\eqref{smith_bivariate_pickands}, we see that it is necessary to set $\delta_{ij} = 2 /a_{\{i,j\}}$, $\{i,j\} \subset D$, for $A_\delta$ to be the Pickands dependence function of $(Z_\th(\vec x_1),\dots,Z_\th(\vec x_d))$. Writing $A_\delta = A_\th$, it follows that the extremal coefficient of the sites in $\{\vec x_i : i \in D\}$ for this model is
$$
\xi_D = d A_\th(1/d, \dots, 1/d) = \sum_{j=1}^d {\Phi^{(d-1)}_{{\bm \Gamma}_j} \left( \frac{a_{\{i,j\}}}{2} : i \in D \setminus \{j\} \right)}.
$$
More generally, for any $B \subset D$ with $|B| \geq 2$,  
\begin{equation}
\label{extremal_Smith}
\xi_B = |B| A_\th(\vec w_b) = \sum_{j \in B} {\Phi^{(|B|-1)}_{{\bm \Gamma}_{j,B}} \left( \frac{a_{\{i,j\}}}{2} : i \in B \setminus \{j\} \right)},
\end{equation}
where ${\bm \Gamma}_{j,B}$ is the $(|B|-1) \times (|B|-1)$ matrix obtained from ${\bm \Gamma}_j$ be removing rows and columns whose index is not in $B$. The previous expression can be computed provided one can compute the c.d.f.\ of the multivariate normal distribution. In \textsf{R}, this can be done using the excellent {\tt mvtnorm} package \citep{mvtnorm}.

Hence, once~\eqref{composite} has been maximized, the resulting estimate can be plugged into~\eqref{extremal_Smith} to obtain the estimate of $\xi_B$ under the null given by~\eqref{estimator_theta_null}.

\subsubsection{The case of the Schlather model} 
\label{Schlather}

For the Schlather model, the dependence is controlled by the correlation function $\rho$, i.e., $\rho = \rho_\th$, and, as for most max-stable processes, the expression of $C_\th$ is available in closed form only in dimension two. Starting from~\eqref{schlather_bivariate}, one obtains that the Pickands dependence function of the random vector $(Z_\th(\vec x_j),Z_\th(\vec x_k))$, for a pair of sites $\{\vec x_j, \vec x_k\}$, is given by  
\begin{equation}
\label{schlather_bivariate_pickands}
%A_\th(w) = \frac{1}{2} \left( w_1 + w_2 \right) \left( 1 + \left[1 - \frac{2 \{ \rho(\|x_j-x_k\|) + 1\}w_1w_2}{(w_1 + w_2)^2} \right]^{1/2} \right), \qquad w \in \Delta_1.
A_\th(\vec w) = \frac{1}{2} \left( 1 + \left[1 - 2 \{ \rho(\vec x_j- \vec x_k) + 1\}w_1w_2 \right]^{1/2} \right), \qquad \vec w \in \Delta_1.
\end{equation}
The previous expression can be used to obtain the expression of the p.d.f.\ $c^{(j,k)}_\th$ needed in~\eqref{composite}, and the expression of the extremal coefficient of $\xi_{\{j,k\}}$, which is simply
\begin{equation}
\label{extremal_Schlather}
\xi_{\{j,k\}} = 1 + \left[\frac{1- \rho(\vec x_j- \vec x_k)}{2} \right]^{1/2}.
\end{equation}
Because of the unavailability of the expression of the Pickands dependence function in dimension three or greater, we do not have a closed form expression for $\xi_B = |B| A_\th(\vec w_B)$ under the Schlather model for $B \subset D$, $|B| \geq 3$. However, from the work of \citet{Sch02}, we know how to generate a random sample from $C_\th$, which enables us to use the estimator given in~\eqref{estimator_theta_null_2}.

\subsection{Asymptotic distribution of the test statistics under the null}
\label{asymptotic}

For a subset $B \subset D$ with $|B| \geq 2$, let $\hat \xi_{B,n}$ denote one of the three nonparametric estimators of $\xi_B$ defined in~\eqref{extremal_estimators}, and recall that $\xi_{B,\hat \th_n}$ and $\hat \xi_{B,\hat \th_n,m}$ are the estimators under $H_0$ of $\xi_B = \xi_{B,\th_0}$ defined in~\eqref{estimator_theta_null} and~\eqref{estimator_theta_null_2}, respectively. Finally, let $\dot \xi_{B,\th}$ be the gradient of $\xi_{B,\th}$ with respect to $\th$.

The following proposition is a consequence of the delta method and the continuous mapping theorem.

\begin{prop}
\label{weak_limit_1}
Assume that $H_0$ holds, that $\sqrt{n} \left( \hat \xi_{B,n} - \xi_{B,\th_0}, \hat \th_n - \th_0 \right)$ converges in distribution to $(\Lambda_B,\bm \Theta)$ and that $\th \mapsto \xi_{B,\th}$ is differentiable at $\th_0$. Then, the test statistic $S_{B,n} =  | \sqrt{n} (\hat \xi_{B,n} - \xi_{B,\hat \th_n}) |$ converges in distribution to $| \Lambda_B - \dot \xi_{B,\th_0}^\top \bm \Theta |$. 
\end{prop}

The convergence in distribution of $\sqrt{n} ( \hat \xi_{B,n} - \xi_{B,\th_0} )$ occurs if Conditions 2.1 and 4.1 of \citet{Seg12} are satisfied. These smoothness conditions concern the first and second-order partial derivatives of~$C_{\th_0}$. If they are satisfied, the limiting random variable can be expressed in terms of the weak limit of the empirical process $\sqrt{n} (\hat A_n - A_{\th_0})$ established in Theorem~1 of \citet{GudSeg12} \citep[see also][Theorem 3.2]{GenSeg09}, which in turn depends on the weak limit of the empirical copula process \citep[see e.g.][]{Seg12}.

In dimension three or higher, the verification of Conditions 2.1 and 4.1 of \citet{Seg12} seems impossible for the Schlather and geometric Gaussian models as a closed form expression of $C_{\th_0}$ is not available in those cases, and appears very tedious for the Smith model. In dimension two, \citet{Seg12} showed that the aforementioned smoothness conditions are satisfied if the function $f(t) = A_{\th_0}(t,1-t)$, $t \in [0,1]$, is twice continuously differentiable on $(0,1)$, and if $\sup_{t \in (0,1)} \{t(1-t) f''(t)\} < \infty$. The latter conditions on $f$ appear to hold for the Smith model, the Schlather and the geometric Gaussian models.

Regularity conditions under which $\sqrt{n} \left( \hat \xi_{B,n} - \xi_{B,\th_0}, \hat \th_n - \th_0 \right)$ converges in distribution still need to be established. A preliminary task would be to obtain regularity conditions for the asymptotic normality of the maximum pairwise pseudo-likelihood estimator. Such regularity conditions are investigated  in \citet{GenGhoRiv95} for the maximum pseudo-likelihood estimator and in \citet{PadRibSis10} for the maximum pairwise likelihood estimator.

Let us now state an analogue of Proposition~\ref{weak_limit_1} for the test statistic $S_{B,n,m} = | \sqrt{n} (\hat \xi_{B,n} - \hat \xi_{B,\hat \th_n,m}) |$, $B \subset D$, $|B| \geq 2$. For any $\th \in \OO$, recall that $c_\th$ is the density associated with $C_\th$, and denote by $\dot c_\th$ and $\dot C_\th$ the gradients with respect to $\th$ of $c_\th$ and $C_\th$, respectively. The following technical conditions are considered:
\begin{enumerate}[({A}1)]
\item The family of copulas $\{C_\th : \th \in \OO\}$ satisfies the regularity conditions stated in Definition~1 of \cite{GenRem08} \citep[see also][Appendix~B~(a)]{GenKojNesYan11} as well as Conditions 2.1 and 4.1 of \citet{Seg12}.
\item For every $\th \in \OO$, $\vth \mapsto \xi_{B,\vth}$ is differentiable at $\th$.
\item For every $\th \in \OO$ and every $\vec w \in \Delta_{d-1}$, there exists a neighborhood $\NN$ of $\th$ and Lebesgue integrable functions $h,g:(0,1) \to \R$ such that
$$
\sup_{\vth \in \NN} \left\| \frac{\dot C_\vth(\vec u^{\vec w})}{u} \right\| \leq h(u) \qquad \mbox{and} \qquad \sup_{\vth \in \NN} \left\| \frac{\dot C_\vth(\vec u^{\vec w})}{u \log(u)} \right\| \leq g(u) \qquad \forall \, u \in (0,1),
$$
where $\vec u^{\vec w} = (u^{w_1},\dots,u^{w_d})$. 
\end{enumerate}
Finally, let $(U_{1,1},\dots,U_{1,d}),\dots,(U_{n,1},\dots,U_{n,d})$ be the unobservable random sample obtained from the available one by $U_{i,j} = F_j(X_{i,j})$, $i \in \{1,\dots,n\}$, $j \in \{1,\dots,d\}$. The following result is then essentially a consequence of Theorem~2 of \cite{GenRem08}.

\begin{prop}
\label{weak_limit_2}
Assume that (A1)--(A3) and $H_0$ hold, and that 
\begin{equation}
\label{assumptions}
 \left( \sqrt{n} (\hat \xi_{B,n} - \xi_{B,\th_0}) ,  \sqrt{n} (\hat \th_n - \th_0), \frac{1}{\sqrt{n}} \sum_{i=1}^n \frac{\dot c_{\th_0}(U_{i,1},\dots,U_{i,d})}{c_{\th_0}(U_{i,1},\dots,U_{i,d})} \right) \leadsto (\Lambda_B,\bm \Theta,\Wb),
\end{equation}
%with $\Ex(\th\Wb) = I$ 
where the arrow $\leadsto$ denotes convergence in distribution.  
Then, the test statistic $S_{B,n,m} =  | \sqrt{n} (\hat \xi_{B,n} - \hat \xi_{B,\hat \th_n,m}) |$, with $m = \lfloor \gamma n \rfloor$, converges in distribution to $| \Lambda_B - \gamma^{-1/2} \Lambda_B' - \dot \xi_{B,\th_0}^\top \bm \Theta |$, where $\Lambda_B'$ is an independent copy of $\Lambda_B$. 
\end{prop}

From the previous proposition, we see that the limiting distribution of $S_{B,n,m}$ under $H_0$ contains the additional term $\gamma^{-1/2} \Lambda_B'$ compared with that of $S_{B,n}$ given in Proposition~\ref{weak_limit_1}. The influence of that term can be made arbitrarily small by taking $\gamma$ sufficiently large.

\subsection{The goodness-of-fit procedures}

The weak limits established in Propositions~\ref{weak_limit_1} and~\ref{weak_limit_2} are unwieldy and cannot be used to compute asymptotic $p$-values for the test statistics. For a subset $B \subset D$ with $|B| \geq 2$, approximate $p$-values for $S_{B,n}$ and $S_{B,n,m}$ can however be obtained using a {\em one-level} and a {\em two-level parametric bootstrap}, respectively. These two procedures are described in the forthcoming two subsections. In the rank-based context under consideration, these resampling techniques were studied by \citet{GenRem08}, who derived technical conditions for their asymptotic validity. When adapted to the current setting, these conditions are almost exactly those used in Proposition~\ref{weak_limit_2}: If (A1)--(A3) hold and if, under $H_0$, \eqref{assumptions} holds with $\Ex(\bm \Theta \Wb^\top) = \bm I$, where $\bm I$ is the $p \times p$ identity matrix, then the one- and two-level parametric bootstrap procedures given below are asymptotically valid in the sense of Theorems~1 and~2 of \cite{GenRem08}. As a consequence, under the validity conditions and $H_0$, each test statistic and its bootstrap replicates converge jointly in distribution to independent copies of the same limit. As already mentioned, related validity conditions can be found in \citet[Appendix~B]{GenKojNesYan11}. 

%In the case of bivariate goodness-of-fit tests based on statistics derived from empirical processes of the form $\sqrt{n} (\hat A_n - A_{\hat \th_n})$,  validity conditions for the one- and two-level parametric bootstraps are listed in \citet[Appendix~B]{GenKojNesYan11}. These conditions can be extended to the $d$-dimensional case and are very similar to the conditions listed in the previous subsection.

As we continue, $N$ and $m$ are large integers and correspond to the number of bootstrap replicates and to the size of the second-level bootstrap sample, respectively.

\subsubsection{A one-level parametric bootstrap for the test based on $S_{B,n}$}
\label{onelevel}

\begin{enumerate}
\item Compute $\hat \th_n$ as a maximizer of~\eqref{composite} and $\hat \xi_{B,n}$ from the available sample. 

\item Compute the test statistic $S_{B,n} = | \sqrt{n} (\hat \xi_{B,n} - \xi_{B,\hat \th_n}) |$.

\item For every $k \in \{1,\dots,N\}$, repeat the following steps:

  \begin{enumerate}[(a)]

  \item Generate a random sample $(U_{1,1}^{(k)},\dots,U_{1,d}^{(k)}),\dots,(U_{n,1}^{(k)},\dots,U_{n,d}^{(k)})$ from $C_{\hat \th_n}$ and compute the corresponding pseudo-observations.

  \item Let $\hat \th_n^{(k)}$ and $\hat \xi_{B,n}^{(k)}$ be the versions of $\hat \th_n$ and $\hat \xi_{B,n}$ computed from the pseudo-observations obtained in Step~(a).
    
  \item Form an approximate realization of $S_{B,n}$ under the null as $S_{B,n}^{(k)} = | \sqrt{n} (\hat \xi_{B,n}^{(k)} - \xi_{B,\hat \th_n^{(k)}}) |$.
\end{enumerate}
\item An approximate $p$-value for $S_{B,n}$ is given by $N^{-1} \sum_{k=1}^N \1(S_{B,n}^{(k)} \geq S_{B,n})$.
\end{enumerate}

\subsubsection{A two-level parametric bootstrap for the test based on $S_{B,n,m}$}
\label{twolevel}

\begin{enumerate}
\item Compute $\hat \th_n$ as a maximizer of~\eqref{composite} and $\hat \xi_{B,n}$ from the available sample.

\item Generate a random sample $(V_{1,1},\dots,V_{1,d}),\dots,(V_{m,1},\dots,V_{m,d})$ from $C_{\hat \th_n}$, and compute $\hat \xi_{B,\hat \th_n,m}$ from the corresponding pseudo-observations using~\eqref{estimator_theta_null_2}.

\item Compute the test statistic $S_{B,n,m} = | \sqrt{n} (\hat \xi_{B,n} - \hat \xi_{B,\hat \th_n,m}) |$.

\item For every $k \in \{1,\dots,N\}$, repeat the following steps:
  \begin{enumerate}[(a)]
  \item Generate a random sample $(U_{1,1}^{(k)},\dots,U_{1,d}^{(k)}),\dots,(U_{n,1}^{(k)},\dots,U_{n,d}^{(k)})$ from $C_{\hat \th_n}$ and compute the corresponding pseudo-observations.
  \item Let $\hat \th_n^{(k)}$ and $\hat \xi_{B,n}^{(k)}$ be the versions of $\hat \th_n$ and $\hat \xi_{B,n}$ computed from the pseudo-observations obtained in Step~(a).
\item Generate a random sample $(V_{1,1}^{(k)},\dots,V_{1,d}^{(k)}),\dots,(V_{m,1}^{(k)},\dots,V_{m,d}^{(k)})$ from $C_{\hat \th_n^{(k)}}$, and compute $\hat \xi_{B,\hat \th_n^{(k)},m}^{(k)}$ from the corresponding pseudo-observations using~\eqref{estimator_theta_null_2}.
    
  \item Form an approximate realization of $S_{B,n,m}$ under the null as $S_{B,n,m}^{(k)} = | \sqrt{n} (\hat \xi_{B,n}^{(k)} - \hat \xi_{B,\hat \th_n^{(k)},m}^{(k)} ) |$.
\end{enumerate}
\item An approximate $p$-value for $S_{B,n}$ is given by $N^{-1} \sum_{k=1}^N \1(S_{B,n,m}^{(k)} \geq S_{B,n})$.
\end{enumerate}

\section{Monte Carlo experiment}
\label{simulations}

As already mentioned in Section~\ref{description}, test statistics of the following form were considered in the simulations: 
\begin{multline}
\label{statistics}
E_{n,2}^{[\alpha]} = \sum_{B \subset D \atop |B| = 2} \{S_{B,n}\}^\alpha, \qquad E_{n,3}^{[\alpha]} = \sum_{B \subset D \atop |B| = 3} \{S_{B,n}\}^\alpha, \qquad S_{D,n}, \qquad E_{n,2,3,d}^{[\alpha]} = \sum_{B \subset D \atop |B| = 2,3,d} \{S_{B,n}\}^\alpha, \\ \mbox{and} \qquad P_{n,2,3,d}^{[\alpha]} = \sum_{B \subset D \atop |B| = 2,3,d} \{S_{B,n} / |B|\}^\alpha, \qquad \alpha \in \{1,2\},
\end{multline} 
where $S_{B,n}$ is defined in~\eqref{generic}. The first type of test statistic can be seen as focusing on the difference between a nonparametric and a parametric estimator of the Pickands dependence function on the boundary of the unit simplex $\Delta_{d-1}$, while the third one considers this difference in the center of $\Delta_{d-1}$. The difference between $E_{n,2,3,d}^{[\alpha]}$ and $P_{n,2,3,d}^{[\alpha]}$ is that the former sums differences of extremal coefficients while the latter sums differences of Pickands dependence functions. By setting $\alpha$ to 2, one obtains Cram\'er--von Mises-like statistics. Three versions of each test statistic can be computed, depending on which of the three nonparametric estimators of the extremal coefficients defined in~\eqref{extremal_estimators} is used. Recall that the latter can be the Pickands estimator, the Hall-Tajvidi estimator or the Cap\'era\`a--Foug\`eres--Genest estimator. %The abbreviations P, HT and CFG, respectively, will be used in the sequel.

\begin{figure}[tbp]
  \centering
  \includegraphics[width=\textwidth]{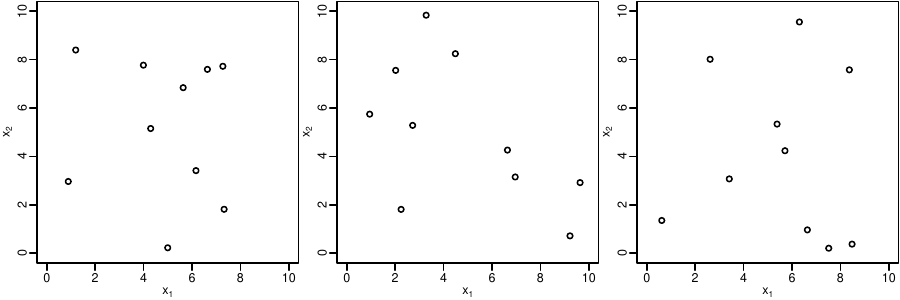}
  \caption{The three different sets of $d=10$ sites used in the simulations.}
  \label{fig:sites}
\end{figure}

The finite-sample performance of the tests was investigated in a computationally intensive Monte Carlo experiment using $[0, 10]^2$ as study region and $d=10$ sites. The factors of the experiment are the locations of the sites, the data generating model, the hypothesized model, the strength of the spatial dependence and the sample size $n$ (typically corresponding to the number of years in a real dataset). To avoid increasing an already very high computational burden, only isotropic models with one real parameter $\theta > 0$ were considered. The first model, abbreviated by Sm--Iso, was obtained by parametrizing the covariance matrix $\bm \Sigma$ in the Smith model as ${\bm \Sigma} = \theta \bm I_2$, where $\bm I_2$ is the $2 \times 2$ identity matrix. The second model, abbreviated as Sc--Exp, was obtained by choosing the correlation function $\rho$ parametrizing the Schlather model to be of the exponential type with range parameter $\theta$, i.e., 
\begin{equation}
\label{rhoexp}
\rho_{\mathrm{exp}}(\vec x) = \exp(- \|\vec x\| / \theta), \qquad \vec x \in \R^2.
\end{equation}
The last model, a particular geometric Gaussian model abbreviated as GG--Exp, was obtained by fixing the parameter $\sigma^2$ in~\eqref{geogaussian} to 8 and by using the exponential correlation function given by~\eqref{rhoexp}. For each of the three models, three values of $\theta$ were considered for random number generation. They were chosen so that the bivariate extremal coefficient $\xi_{\{i,j\}}$ of two fictitious sites $\vec x_i$ and $\vec x_j$ equals 1.5 when the distance between $\vec x_i$ and $\vec x_j$ equals 1, 5, and 10, respectively. The latter distance will be denoted by $d_{1.5}$ as we continue. The sample size $n$ was taken in $\{50,100,200\}$. To investigate the influence of the locations of the $d=10$ sites, three different sets of sites were generated. These are represented in Figure~\ref{fig:sites}. A larger number of site configurations was not considered for computational reasons. 

Samples from the Sm-Iso, Sc-Exp and GG-Exp models were generated using the excellent \texttt{SpatialExtremes} \textsf{R} package \citep{SpatialExtremes}. Note that $\sigma^2$ was set to 8 in the model GG--Exp because random number generation for the geometric Gaussian model in \texttt{SpatialExtremes} is apparently unreliable when $\sigma^2 > 10$. For each set of sites, each of the three models and each value of $\theta$, 1000 samples were produced.  For each generated sample, the goodness of fit of the models Sm-Iso, Sc--Exp and GG-Exp was tested. The bootstrap sample size $N$ was set to 1000 and all tests were carried out at the 5\% significance level. For the tests based on $E_{n,2}^{[\alpha]}$ defined in~\eqref{statistics}, the one-level parametric bootstrap of Section~\ref{onelevel} was used as a closed-form expression of the bivariate extremal coefficient is available for all three models considered in the simulations (as for most models). To avoid the use of the more costly two-level parametric bootstrap of Section~\ref{twolevel} for the tests based on the other statistics given in~\eqref{statistics}, we ``precomputed'' reasonnably accurate approximations of the mappings $\theta \mapsto \xi_{B,\theta}$ for all three sets of sites displayed in Figure~\ref{fig:sites}, all three models and all $B \subset D$, $|B| \in \{3,d\}$. This was done using the procedure described in detail in Appendix~\ref{compcost} and enabled us to save a lot of computing time. Note that the use of the two-level parametric bootstrap will be presented in the illustration of Section~\ref{illustration}, where it will be also compared with the results of the one-level parametric bootstrap when assessing the fit of the Smith model.

The obtained rejection percentages of $H_0$ for the $d=10$ sites represented in the left (resp.\ middle, right) plot of Figure~\ref{fig:sites} are given in Table~\ref{tab:sm1} (resp.\ \ref{tab:sm2}, \ref{tab:sm3}). The values of $\theta$ used for data generation are given in the third column of the tables, while the second column recalls the corresponding value of $d_{1.5}$ (the distance between two sites for their extremal coefficient to be equal to 1.5). The tables only report the rejection percentages for the tests based on the statistics in~\eqref{statistics} when based on the CFG estimator of the unknown Pickands dependence function. Indeed, with a very few exceptions, the use of the CFG estimator led to substantially more powerful tests. The results for the test statistics with $\alpha=2$ are not reported as the corresponding tests did not appear more powerful than those with~$\alpha=1$. A final general remark is that the results appear to be qualitatively identical for the three sets of sites.

\begin{sidewaystable}[tbp]
\addtolength{\tabcolsep}{-2pt}
\begin{center}
\caption{Percentage of rejection of $H_0$ computed from 1000 samples of size $n$ generated from the models Sm--Iso, Sc--Exp and GG--Exp with parameter value $\theta$ for the $d=10$ sites represented in the left plot of Figure~\ref{fig:sites}.}
\label{tab:sm1}
\begin{tabular}{cccc rrrrr rrrrr rrrrr}
\toprule
Model & $d_{1.5}$ & $\theta$ & $n$ & \multicolumn{5}{c}{$H_0$ : Sm--Iso} & \multicolumn{5}{c}{$H_0$ : Sc--Exp} & \multicolumn{5}{c}{$H_0$ : GG--Exp}\\
\cmidrule(lr){5-9}\cmidrule(lr){10-14}\cmidrule(lr){15-19}
  & & & & $S_{D,n}$ & $E_{n,3}^{[1]}$ & $E_{n,2}^{[1]}$ &  $E_{n,2,3,d}^{[1]}$ &  $P_{n,2,3,d}^{[1]}$ &  $S_{D,n}$ & $E_{n,3}^{[1]}$ & $E_{n,2}^{[1]}$ &  $E_{n,2,3,d}^{[1]}$ &  $P_{n,2,3,d}^{[1]} $  &  $S_{D,n}$ & $E_{n,3}^{[1]}$ & $E_{n,2}^{[1]}$ &  $E_{n,2,3,d}^{[1]}$ &  $P_{n,2,3,d}^{[1]}$ \\
%& & & & All & Tps & Prs & Cmb & CmbS & All & Tps & Prs & Cmb & CmbS & All & Tps & Prs & Cmb & CmbS \\
\midrule
Sm--Iso & 1 & 0.550 & 50 & {\it  11.2} & {\it  12.4} & {\it  11.8} & {\it  12.2} & {\it  12.4} & 100.0 & 100.0 &  99.9 & 100.0 & 100.0 &   1.9 &   0.6 &   4.1 &   0.8 &   1.2 \\ 
   &  &  & 100 & {\it   5.9} & {\it   5.6} & {\it   5.0} & {\it   5.7} & {\it   5.7} & 100.0 & 100.0 & 100.0 & 100.0 & 100.0 &   0.8 &   7.7 &  16.1 &   6.9 &   8.8 \\ 
   &  &  & 200 &  {\it 5.1} & {\it 5.8} & {\it 4.9} & {\it 5.5} & {\it 5.2} &  &  &  &  &  &  &  &  &  &  \\ 
   & 5 & 13.74 & 50 & {\it   4.9} & {\it   5.5} & {\it   4.7} & {\it   5.3} & {\it   5.2} &   5.3 &  45.3 &  70.4 &  52.1 &  55.4 &   4.6 &  80.5 &  96.7 &  86.8 &  89.5 \\ 
   &  &  & 100 & {\it   6.5} & {\it   7.3} & {\it   5.7} & {\it   7.0} & {\it   6.6} &   7.4 &  99.3 & 100.0 &  99.9 & 100.0 &   1.4 & 100.0 & 100.0 & 100.0 & 100.0 \\ 
   & 10 & 54.95 & 50 & {\it   2.1} & {\it   2.3} & {\it   2.5} & {\it   2.3} & {\it   2.2} &   0.2 &   5.8 &  30.3 &   8.4 &  10.2 &   1.2 &  85.6 &  93.6 &  89.1 &  90.0 \\ 
   &  &  & 100 & {\it   3.4} & {\it   2.9} & {\it   3.7} & {\it   3.0} & {\it   3.1} &   1.1 &  89.1 & 100.0 &  96.7 &  97.8 &   1.3 & 100.0 & 100.0 & 100.0 & 100.0 \\ 
   [2ex]\\
Sc--Exp & 1 & 1.443 & 50 & 100.0 & 100.0 & 100.0 & 100.0 & 100.0 & {\it   6.1} & {\it   8.2} & {\it   6.9} & {\it   8.0} & {\it   7.8} & 100.0 & 100.0 &  98.5 & 100.0 & 100.0 \\ 
   &  &  & 100 & 100.0 & 100.0 & 100.0 & 100.0 & 100.0 & {\it   5.2} & {\it   7.6} & {\it   7.9} & {\it   7.8} & {\it   7.7} & 100.0 & 100.0 & 100.0 & 100.0 & 100.0 \\ 
   &  &  & 200 &   &  &  &  &  & {\it 5.3} & {\it 7.0} & {\it 6.3} & {\it 7.1} & {\it 6.9} &  &  &  &  &  \\ 
   & 5 & 7.213 & 50 & 100.0 & 100.0 & 100.0 & 100.0 & 100.0 & {\it   4.7} & {\it   3.5} & {\it   2.7} & {\it   3.3} & {\it   3.1} &  97.7 &  98.1 &  83.9 &  98.0 &  97.2 \\ 
   &  &  & 100 &  99.6 & 100.0 & 100.0 & 100.0 & 100.0 & {\it   4.6} & {\it   4.6} & {\it   4.7} & {\it   4.9} & {\it   4.6} & 100.0 & 100.0 &  99.8 & 100.0 & 100.0 \\ 
   & 10 & 14.43 & 50 &  99.7 & 100.0 & 100.0 & 100.0 & 100.0 & {\it   5.2} & {\it   4.1} & {\it   3.4} & {\it   3.6} & {\it   3.6} &  91.5 &  92.7 &  88.5 &  92.9 &  92.9 \\ 
   &  &  & 100 &  99.0 & 100.0 & 100.0 & 100.0 & 100.0 & {\it   4.1} & {\it   2.9} & {\it   3.0} & {\it   2.5} & {\it   2.8} &  99.6 & 100.0 & 100.0 & 100.0 & 100.0 \\ 
   [2ex]\\
GG--Exp & 1 & 8.282 & 50 &  76.5 &  99.2 &  99.2 &  99.3 &  99.3 &  93.8 &  50.9 &  30.2 &  51.7 &  49.2 & {\it   9.7} & {\it   7.8} & {\it   8.9} & {\it   8.1} & {\it   8.1} \\ 
   &  &  & 100 &  92.7 &  99.9 & 100.0 &  99.9 &  99.9 & 100.0 &  88.3 &  73.9 &  90.2 &  87.3 & {\it   8.7} & {\it   6.7} & {\it   5.9} & {\it   7.3} & {\it   7.1} \\ 
   &  &  & 200 &   &  &  &  &  &  &  &  &  &  & {\it 5.8} & {\it 8.1} & {\it 7.7} & {\it 8.1} & {\it 8.0} \\ 
   & 5 & 41.41 & 50 &  69.8 & 100.0 & 100.0 & 100.0 & 100.0 &   7.8 &   4.3 &   1.4 &   3.7 &   3.1 & {\it   5.2} & {\it   3.6} & {\it   3.8} & {\it   3.9} & {\it   4.0} \\ 
   &  &  & 100 &  89.1 & 100.0 & 100.0 & 100.0 & 100.0 &  15.9 &  19.7 &  27.1 &  20.3 &  21.7 & {\it   4.9} & {\it   4.4} & {\it   5.4} & {\it   4.4} & {\it   4.5} \\ 
   & 10 & 82.82 & 50 &  80.4 & 100.0 & 100.0 & 100.0 & 100.0 &   3.1 &  19.0 &   2.9 &  15.9 &  14.5 & {\it   3.8} & {\it   3.0} & {\it   3.6} & {\it   3.3} & {\it   3.4} \\ 
   &  &  & 100 &  96.7 & 100.0 & 100.0 & 100.0 & 100.0 &   2.3 &  75.2 &  78.9 &  76.4 &  77.9 & {\it   4.5} & {\it   4.8} & {\it   5.0} & {\it   4.7} & {\it   4.4} \\ 
\bottomrule
\end{tabular}
\end{center}
\end{sidewaystable}

\begin{sidewaystable}[tbp]
\addtolength{\tabcolsep}{-2pt}
\begin{center}
\caption{Percentage of rejection of $H_0$ computed from 1000 samples of size $n$ generated from the models Sm--Iso, Sc--Exp and GG--Exp with parameter value $\theta$ for the $d=10$ sites represented in the middle plot of Figure~\ref{fig:sites}.}
\label{tab:sm2}
\begin{tabular}{cccc rrrrr rrrrr rrrrr}
\toprule
Model & $d_{1.5}$ & $\theta$ & $n$ & \multicolumn{5}{c}{$H_0$ : Sm--Iso} & \multicolumn{5}{c}{$H_0$ : Sc--Exp} & \multicolumn{5}{c}{$H_0$ : GG--Exp}\\
\cmidrule(lr){5-9}\cmidrule(lr){10-14}\cmidrule(lr){15-19}
  & & & & $S_{D,n}$ & $E_{n,3}^{[1]}$ & $E_{n,2}^{[1]}$ &  $E_{n,2,3,d}^{[1]}$ &  $P_{n,2,3,d}^{[1]}$ &  $S_{D,n}$ & $E_{n,3}^{[1]}$ & $E_{n,2}^{[1]}$ &  $E_{n,2,3,d}^{[1]}$ &  $P_{n,2,3,d}^{[1]} $  &  $S_{D,n}$ & $E_{n,3}^{[1]}$ & $E_{n,2}^{[1]}$ &  $E_{n,2,3,d}^{[1]}$ &  $P_{n,2,3,d}^{[1]}$ \\
% & & & & All & Tps & Prs & Cmb & CmbS & All & Tps & Prs & Cmb & CmbS & All & Tps & Prs & Cmb & CmbS \\
\midrule
Sm--Iso & 1 & 0.550 & 50 & {\it  11.0} & {\it  11.1} & {\it   9.0} & {\it  11.3} & {\it  11.2} & 100.0 & 100.0 &  99.9 & 100.0 & 100.0 &   3.0 &   0.6 &   1.8 &   0.6 &   0.6 \\ 
   &  &  & 100 & {\it   7.0} & {\it   5.7} & {\it   5.0} & {\it   5.5} & {\it   5.3} & 100.0 & 100.0 & 100.0 & 100.0 & 100.0 &   1.1 &   0.4 &   4.6 &   0.3 &   0.6 \\ 
   &  &  & 200 & {\it 6.7} & {\it 5.5} & {\it 5.0} & {\it 5.8} & {\it 5.5} &  &  &  &  &  &  &  &  &  &  \\ 
   & 5 & 13.74 & 50 & {\it   4.8} & {\it   6.5} & {\it   4.8} & {\it   6.5} & {\it   6.2} &   3.2 &  41.4 &  80.6 &  50.6 &  55.7 &  15.6 &  68.0 &  95.6 &  81.2 &  84.4 \\ 
   &  &  & 100 & {\it   4.6} & {\it   5.7} & {\it   5.0} & {\it   5.9} & {\it   5.9} &   4.3 &  97.1 & 100.0 &  99.3 &  99.5 &  18.4 &  99.4 & 100.0 & 100.0 & 100.0 \\ 
   & 10 & 54.95 & 50 & {\it   3.2} & {\it   3.0} & {\it   2.7} & {\it   2.9} & {\it   2.8} &   5.5 &   9.0 &  48.8 &  16.2 &  19.6 &   6.3 &  81.8 &  95.3 &  88.2 &  89.0 \\ 
   &  &  & 100 & {\it   3.2} & {\it   3.5} & {\it   3.3} & {\it   3.3} & {\it   3.3} &  25.9 &  94.4 & 100.0 &  98.8 &  99.5 &  12.9 &  99.9 & 100.0 & 100.0 & 100.0 \\ 
   [2ex]\\
Sc--Exp & 1 & 1.443 & 50 & 100.0 & 100.0 & 100.0 & 100.0 & 100.0 & {\it   4.2} & {\it   9.6} & {\it   7.0} & {\it   9.3} & {\it   9.3} &  99.9 &  99.8 &  98.5 &  99.9 &  99.8 \\ 
   &  &  & 100 & 100.0 & 100.0 & 100.0 & 100.0 & 100.0 & {\it   3.6} & {\it   6.3} & {\it   5.7} & {\it   6.5} & {\it   6.5} & 100.0 & 100.0 & 100.0 & 100.0 & 100.0 \\ 
   &  &  &  200 &  &  &  &  &  & {\it 2.5} & {\it 5.6} & {\it 6.3} & {\it 5.4} & {\it 5.3} &  &  &  &  &  \\ 
   & 5 & 7.213 & 50 & 100.0 & 100.0 & 100.0 & 100.0 & 100.0 & {\it   4.9} & {\it   4.8} & {\it   4.2} & {\it   4.6} & {\it   4.6} &  97.6 &  97.2 &  83.5 &  97.4 &  96.3 \\ 
   &  &  & 100 & 100.0 & 100.0 & 100.0 & 100.0 & 100.0 & {\it   5.9} & {\it   4.6} & {\it   4.5} & {\it   4.6} & {\it   4.8} & 100.0 & 100.0 &  99.4 & 100.0 & 100.0 \\ 
   & 10 & 14.43 & 50 &  99.8 & 100.0 & 100.0 & 100.0 & 100.0 & {\it   4.8} & {\it   4.3} & {\it   3.3} & {\it   3.5} & {\it   3.3} &  94.5 &  94.3 &  89.2 &  94.9 &  94.4 \\ 
   &  &  & 100 & 100.0 & 100.0 & 100.0 & 100.0 & 100.0 & {\it   4.3} & {\it   4.7} & {\it   4.6} & {\it   4.4} & {\it   4.0} &  99.6 &  99.9 &  99.9 &  99.9 &  99.9 \\ 
   [2ex]\\
GG--Exp & 1 & 8.282 & 50 &  44.3 &  92.0 &  91.7 &  92.4 &  92.5 &  84.0 &  24.2 &  20.9 &  27.5 &  24.6 & {\it  11.1} & {\it   9.3} & {\it   8.0} & {\it   8.8} & {\it   8.9} \\ 
   &  &  & 100 &  58.5 &  99.6 &  99.7 &  99.6 &  99.6 &  99.7 &  76.9 &  82.4 &  80.6 &  79.6 & {\it   8.1} & {\it  11.3} & {\it   9.5} & {\it  10.9} & {\it  10.7} \\ 
  &  &  & 200 &  &  &  &  &  &  &  &  &  &  & {\it 6.7} & {\it 7.3} & {\it 7.4} & {\it 7.6} & {\it 7.5} \\  
   & 5 & 41.41 & 50 &  32.7 &  99.6 & 100.0 &  99.6 &  99.6 &  11.1 &   3.0 &   3.0 &   2.7 &   2.5 & {\it   5.3} & {\it   3.9} & {\it   5.2} & {\it   4.5} & {\it   4.6} \\ 
   &  &  & 100 &  43.2 & 100.0 & 100.0 & 100.0 & 100.0 &  19.4 &  10.2 &  21.5 &  12.4 &  13.9 & {\it   3.8} & {\it   4.6} & {\it   5.6} & {\it   4.5} & {\it   4.5} \\ 
   & 10 & 82.82 & 50 &  32.2 & 100.0 & 100.0 & 100.0 & 100.0 &   2.1 &  16.1 &   3.6 &  13.4 &  12.2 & {\it   4.1} & {\it   3.0} & {\it   4.8} & {\it   3.5} & {\it   3.6} \\ 
   &  &  & 100 &  33.9 & 100.0 & 100.0 & 100.0 & 100.0 &   3.1 &  58.8 &  67.4 &  60.5 &  63.0 & {\it   5.1} & {\it   3.9} & {\it   4.7} & {\it   4.2} & {\it   4.6} \\ 
\bottomrule
\end{tabular}
\end{center}
\end{sidewaystable}

\begin{sidewaystable}[tbp]
\addtolength{\tabcolsep}{-2pt}
\begin{center}
\caption{Percentage of rejection of $H_0$ computed from 1000 samples of size $n$ generated from the models Sm--Iso, Sc--Exp and GG--Exp with parameter value $\theta$ for the $d=10$ sites represented in the right plot of Figure~\ref{fig:sites}.}
\label{tab:sm3}
\begin{tabular}{cccc rrrrr rrrrr rrrrr}
\toprule
Model & $d_{1.5}$ & $\theta$ & $n$ & \multicolumn{5}{c}{$H_0$ : Sm--Iso} & \multicolumn{5}{c}{$H_0$ : Sc--Exp} & \multicolumn{5}{c}{$H_0$ : GG--Exp}\\
\cmidrule(lr){5-9}\cmidrule(lr){10-14}\cmidrule(lr){15-19}
  & & & & $S_{D,n}$ & $E_{n,3}^{[1]}$ & $E_{n,2}^{[1]}$ &  $E_{n,2,3,d}^{[1]}$ &  $P_{n,2,3,d}^{[1]}$ &  $S_{D,n}$ & $E_{n,3}^{[1]}$ & $E_{n,2}^{[1]}$ &  $E_{n,2,3,d}^{[1]}$ &  $P_{n,2,3,d}^{[1]} $  &  $S_{D,n}$ & $E_{n,3}^{[1]}$ & $E_{n,2}^{[1]}$ &  $E_{n,2,3,d}^{[1]}$ &  $P_{n,2,3,d}^{[1]}$ \\
%& & & & All & Tps & Prs & Cmb & CmbS & All & Tps & Prs & Cmb & CmbS & All & Tps & Prs & Cmb & CmbS \\
\midrule
Sm--Iso & 1 & 0.550 & 50 & {\it  17.0} & {\it  16.7} & {\it  17.4} & {\it  17.1} & {\it  16.9} & 100.0 & 100.0 & 100.0 & 100.0 & 100.0 &   1.5 &   0.4 &   4.6 &   0.5 &   0.5 \\ 
   &  &  & 100 & {\it   7.0} & {\it   7.8} & {\it   7.9} & {\it   7.5} & {\it   7.7} & 100.0 & 100.0 & 100.0 & 100.0 & 100.0 &   0.2 &   7.2 &  18.3 &   5.8 &   8.1 \\ 
   &  &  & 200 & {\it 5.6} & {\it 4.9} & {\it 4.8} & {\it 5.0} & {\it 5.0} &  &  &  &  &  &  &  &  &  &  \\ 
   & 5 & 13.74 & 50 & {\it   4.8} & {\it   5.1} & {\it   3.9} & {\it   4.8} & {\it   5.0} &   8.7 &  55.8 &  77.7 &  61.7 &  64.3 &  11.1 &  75.6 &  95.9 &  83.5 &  85.2 \\ 
   &  &  & 100 & {\it   5.2} & {\it   6.6} & {\it   6.1} & {\it   6.1} & {\it   6.0} &  22.4 &  99.0 &  99.9 &  99.8 &  99.8 &   9.8 &  99.8 & 100.0 &  99.9 &  99.9 \\ 
   & 10 & 54.95 & 50 & {\it   4.8} & {\it   2.8} & {\it   3.4} & {\it   2.9} & {\it   3.0} &   0.7 &   7.8 &  38.2 &  11.8 &  14.6 &   3.6 &  88.1 &  95.7 &  91.5 &  92.4 \\ 
   &  &  & 100 & {\it   2.8} & {\it   3.4} & {\it   3.7} & {\it   3.3} & {\it   3.2} &   1.8 &  92.5 & 100.0 &  98.0 &  99.0 &   1.3 & 100.0 & 100.0 & 100.0 & 100.0 \\ 
   [2ex]\\
Sc--Exp & 1 & 1.443 & 50 & 100.0 & 100.0 & 100.0 & 100.0 & 100.0 & {\it   6.8} & {\it   9.7} & {\it   8.0} & {\it   9.7} & {\it   9.9} &  99.9 & 100.0 &  99.5 & 100.0 & 100.0 \\ 
   &  &  & 100 & 100.0 & 100.0 & 100.0 & 100.0 & 100.0 & {\it   5.8} & {\it   7.9} & {\it   7.3} & {\it   7.9} & {\it   7.6} & 100.0 & 100.0 & 100.0 & 100.0 & 100.0 \\ 
   &  &  &  200 &  &  &  &  &  & {\it 4.7} & {\it 6.9} & {\it 5.3} & {\it 6.8} & {\it 6.6} &  &  &  &  &  \\ 
   & 5 & 7.213 & 50 & 100.0 & 100.0 & 100.0 & 100.0 & 100.0 & {\it   5.6} & {\it   3.7} & {\it   4.2} & {\it   3.7} & {\it   3.9} &  97.5 &  96.9 &  82.6 &  97.0 &  96.4 \\ 
   &  &  & 100 & 100.0 & 100.0 & 100.0 & 100.0 & 100.0 & {\it   5.0} & {\it   3.4} & {\it   3.8} & {\it   3.2} & {\it   3.3} & 100.0 & 100.0 &  99.8 & 100.0 & 100.0 \\ 
   & 10 & 14.43 & 50 & 100.0 & 100.0 & 100.0 & 100.0 & 100.0 & {\it   5.7} & {\it   3.1} & {\it   2.8} & {\it   2.8} & {\it   2.7} &  96.1 &  93.5 &  87.6 &  94.4 &  94.0 \\ 
   &  &  & 100 & 100.0 & 100.0 & 100.0 & 100.0 & 100.0 & {\it   3.6} & {\it   4.8} & {\it   4.6} & {\it   4.3} & {\it   4.1} &  99.9 & 100.0 & 100.0 & 100.0 & 100.0 \\ 
   [2ex]\\
GG--Exp & 1 & 8.282 & 50 &  78.8 &  98.6 &  98.8 &  98.4 &  98.6 &  96.9 &  64.9 &  37.5 &  66.7 &  62.6 & {\it   9.7} & {\it   9.1} & {\it   8.2} & {\it   9.6} & {\it   8.9} \\ 
   &  &  & 100 &  96.3 & 100.0 & 100.0 & 100.0 & 100.0 & 100.0 &  95.8 &  85.8 &  96.6 &  95.2 & {\it  12.0} & {\it   8.0} & {\it   6.0} & {\it   8.2} & {\it   7.5} \\ 
   &  &  & 200 &  &  &  &  &  &  &  &  &  &  & {\it 5.7} & {\it 6.9} & {\it 7.1} & {\it 6.6} & {\it 7.0} \\    & 5 & 41.41 & 50 &  74.4 &  99.9 & 100.0 &  99.9 & 100.0 &  10.3 &   2.3 &   2.5 &   2.2 &   2.6 & {\it   5.9} & {\it   3.6} & {\it   4.8} & {\it   3.8} & {\it   4.0} \\ 
   &  &  & 100 &  90.9 & 100.0 & 100.0 & 100.0 & 100.0 &  22.6 &   8.7 &  16.8 &   9.7 &  10.2 & {\it   5.0} & {\it   4.0} & {\it   4.2} & {\it   3.5} & {\it   3.6} \\ 
   & 10 & 82.82 & 50 &  81.9 & 100.0 & 100.0 & 100.0 & 100.0 &   2.6 &  16.1 &   2.5 &  12.7 &  11.6 & {\it   3.5} & {\it   3.1} & {\it   3.8} & {\it   3.1} & {\it   3.2} \\ 
   &  &  & 100 &  95.5 & 100.0 & 100.0 & 100.0 & 100.0 &   2.3 &  60.4 &  68.1 &  61.7 &  63.8 & {\it   6.1} & {\it   3.9} & {\it   4.3} & {\it   4.2} & {\it   4.1} \\ 
\bottomrule
\end{tabular}
\end{center}
\end{sidewaystable}

By considering the empirical levels of the tests given in italic in the tables, we see that, overall, the tests seem to hold their level reasonably well for $\theta$ values corresponding to a pairwise extremal coefficient of 1.5 at distance 5 or 10 (i.e., $d_{1.5} \in \{5,10\}$ in the tables). They appear however too liberal when $d_{1.5}=1$, although the agreement with the 5\% nominal level clearly improves when $n$ increases from 50 to 200.

From the first vertical block of the tables, we see that, when assessing the fit of the model Sm--Iso, the tests have overall high power, and that it is the test based on $E_{n,2}^{[1]}$ (resp.\ $S_{D,n}$) that seems the most (resp.\ least) powerful. When testing the fit of the model Sc--Exp, we see, from the second vertical block of the tables, that it is the test based on $E_{n,2}^{[1]}$ that is the most powerful when data are generated from the model Sm--Iso. When GG--Exp is used as data generating model and $d_{1.5} \in \{1,5\}$, the test based on $S_{D,n}$ displays overall the highest rejection rates, while when $d_{1.5}=10$, it is either $E_{n,2}^{[1]}$ or $E_{n,3}^{[1]}$. Finally, the rejection rates reported in Tables~\ref{tab:sm1}-\ref{tab:sm3} suggest that the most powerful tests overall for assessing the fit of the model GG--Exp are $E_{n,2}^{[1]}$ (when data are generated from Sm--Iso) and $S_{D,n}$ (when data are generated from Sc--Exp).

Note that, for $d_{1.5}=1$, in most situations, very high rejection rates are observed when the model Sc--Exp is involved. This is unsurprising since, as already mentioned, the Schlather model cannot model spatial independence. In a somehow related way, we see, from the second horizontal block of the tables that the rejection percentages are very close (if not equal) to 100\% when data are generated from the model Sc--Exp and when the fit of one of the two other models is assessed.

Given the large number of factors influencing the power of the tests, it is not surprising that no test appears uniformly better. From a practical perspective, we suggest to at least consider the tests based on $E_{n,2}^{[1]}$ and $S_{D,n}$ since, having in mind the interpretation of the statistics given below~\eqref{statistics}, these tests can be used to identify on which ``regions'' of $\Delta_{d-1}$ the estimated model does not fit.

\section{Illustration}
\label{illustration}

As an illustration, the tests were applied to the Swiss rainfall data analyzed by \citet{DavPadRib12}. The data consist of summer maximum daily precipitation for the years 1962--2008 at 51 weather stations in the Plateau region of Switzerland. Among the eleven models fitted in \citet{DavPadRib12} to the maxima measured at a subset of 35 stations, we restricted our attention to the best Smith, Schlather and geometric Gaussian models in terms of composite likelihood information criterion (CLIC) \citep[see][Table 5]{DavPadRib12}. We considered in particular the Smith model with anisotropic covariance matrix $\bm \Sigma = (\Sigma_{ij})$ (abbreviated as Sm-Ani in the sequel), the Schlather model with exponential correlation function given by~\eqref{rhoexp} (abbreviated as Sc--Exp), and two geometric Gaussian models with Whittle--Mat\'ern correlation function. The latter correlation function is defined by
$$
\rho_{\mathrm{WM}}(\vec x) = \frac{1}{2^{\kappa - 1} \Gamma(\kappa)} (\|\vec x\| / \theta)^{\kappa} K_{\kappa}(\|\vec x\| / \theta), \qquad \vec x \in \R^2,
$$
where $\kappa > 0$ is a smoothing parameter, $\theta > 0$ is the range parameter, $K_{\kappa}$ is the modified Bessel function of order $\kappa$ and $\Gamma$ is the gamma function. The parameters of the first geometric Gaussian model, denoted by GG--WM1, are $\sigma^2$ (see~\eqref{geogaussian}) and $\kappa$, while $\theta$ is fixed to 700 as in \citet[Table 5]{DavPadRib12}. The only parameter of the second geometric Gaussian model, denoted by GG--WM2, is $\kappa$, $\sigma^2$ and $\theta$ being fixed to 8.571 and 700, respectively. The latter model was introduced based on the results given in Table~\ref{tab:fit} because the fit of the model GG--WM1 could not be assessed. Indeed, as already mentioned, random number generation for the geometric Gaussian model in the \texttt{SpatialExtremes} package is apparently only reliable for $\sigma^2 < 10$, and performing a parametric bootstrap for GG--WM1 turned out to produce estimates of $\sigma^2$ frequently larger than 10. Similarly, the goodness of fit of the Brown--Resnick models considered in \citet{DavPadRib12} was not assessed because we had no access to efficient random number generation in the 2-dimensional case.

\begin{table}[t!]
\centering
\label{tab:fit}
\caption{Summary of the max-stable models fitted to the Swiss rainfall data using the \texttt{SpatialExtremes} \textsf{R} package. }
\begin{tabular}{l r@{(} c  @{)\hspace{10pt}} r@{(} c  @{)\hspace{10pt}} r@{(} c  @{)\hspace{10pt}} r r}
\toprule
Model & $\sigma^2$ & se & $\theta$ & se & $\kappa$ & se & loglik & CLIC \\ 
  \midrule
GG--WM1 & 8.571 & 2.256 & 700 & --- & 0.368 & 0.030 & $-$231488 & 463286 \\ 
GG--WM2   & 8.571 & --- & 700 & --- & 0.368 & 0.011 & $-$231488 & 463180 \\ 

Sc--Exp & --- & --- & 42.004 & 6.643 & --- & --- & $-$232167 & 464563 \\ 

\\[2ex]
Model & $\Sigma_{11}$ & se & $\Sigma_{12}$ & se & $\Sigma_{22}$ & se & loglik & CLIC\\ 
\midrule
Sm-Ani & 351.680 & 6.110 & 37.364 & 4.177 & 312.435 & 12.856 & $-$236437 & 472964 
\\ 
\bottomrule
\end{tabular}
\end{table}

Our model fitting was different from \citet{DavPadRib12} in two aspects: first, we used all 51 sites, including the 16 sites left out for validation in \citet{DavPadRib12}; second, the fitting was based on the maximization of the pairwise log pseudo-likelihood given in~\eqref{composite} thereby avoiding the necessary step of estimating marginal parameters in trend surfaces and the risk of misspecification. This explains why the results of the fitting given in Table~\ref{tab:fit}, although similar, do not coincide with those of \citet{DavPadRib12}. 

As a next step, we assessed the goodness of fit of the models GG--WM2, Sc--Exp and Sm--Ani. For the first two models, the two-level parametric bootstrap of Section~\ref{twolevel} was used to obtain an approximate $p$-value with $N = 1000$ and $m = 2500$. For the third model, both the one- and the two-level parametric bootstraps were used. 

From the plots giving the bivariate extremal coefficients versus site distance under the four fitted models \cite[which are very similar to the plots given in Figure 9 of][]{DavPadRib12}, it appears that the distance at which the bivariate extremal coefficients become equal to 1.5 is somewhere between 30 to 40km. Since the study region is approximately a 70km by 80km rectangle, the spatial dependence in the data seems, up to a scale factor, similar to the spatial dependence corresponding to the settings with $d_{1.5}=5$ in the simulation study reported in Section~\ref{simulations}. We have therefore no reason to believe that the goodness-of-fit tests will be too liberal in the setting under consideration.

\begin{table}[t!]
\centering
\caption{Approximate $p$-values and execution times of the goodness-of-fit tests for the max-stable models fitted to the Swiss rainfall data. The two lines for the model Sm--Ani correspond to the two- and the one-level parametric bootstrap, respectively. The timings are in hours and were obtained on a Linux machine with a 3.4GHz CPU.}
\label{tab:gof}
\begin{tabular}{lrrrrr r}
\toprule
Model & $S_{D,n}$ & $E_{n,3}^{[1]}$ & $E_{n,2}^{[1]}$ & $E_{n,2,3,d}^{[1]}$ & $P_{n,2,3,d}^{[1]}$ & Time (h) \\ 
\midrule
GG--WM2 & 0.206 & 0.114 & 0.050 & 0.111 & 0.112 & 7.6\\ 
Sc--Exp  & 0.001 & 0.804 & 0.330 & 0.773 & 0.784 & 4.1 \\ 
Sm--Ani (2-level) & 0.563 & 0.000 & 0.000 & 0.000 & 0.000 & 4.3 \\ 
Sm--Ani (1-level) & 0.582 & 0.000 & 0.000 & 0.000 & 0.000 & 7.9 \\ 
  \bottomrule
\end{tabular}
\end{table}

Table~\ref{tab:gof} gives the approximate $p$-values of the tests based on the statistics in~\eqref{statistics} with $\alpha=1$ and the Pickands dependence function estimated by the CFG estimator. The two lines for the model Sm--Ani correspond to the two- and the one-level parametric bootstrap, respectively. As expected, the results are similar, but maybe slightly surprisingly, the two-level parametric bootstrap is approximately twice faster. This may be explained by the cost of the evaluation of the multivariate normal c.d.f.\ and the form of the closed-expression of the extremal coefficients under the Smith model; see~\eqref{extremal_Smith}. As many tests are performed, the significance level should be adjusted before interpreting the results. For simplicity, we arbitrarily propose to reason at the 1\% level. From the last two lines of Table~\ref{tab:gof}, we see that the model Sm--Ani is rejected by all the tests except the one based on $S_{D,n}$. In other words, under the Sm--Ani model, we have very strong evidence that the parametric and nonparametric estimates of the Pickands dependence function differ significantly on the boundary of $\Delta_{d-1}$, while there is no evidence of disagreement in the center of $\Delta_{d-1}$. On the contrary, for the model Sc--Exp, there is some evidence of disagrement between the nonparametric and parametric estimates in the center of $\Delta_{d-1}$ only. Finally, we see that the GG--WM2 model was not rejected by any test.

\section*{Acknowledgments}

The computationally intensive simulations reported in Section~\ref{simulations} were carried out on the Beowulf cluster of the Department of Statistics, University of Connecticut. This cluster was partially financed by the NSF grant SCREMS (Scientific Computing Research Environments for the Mathematical Sciences) number 0723557.

\appendix

\section{Proof of the proposition~\ref{weak_limit_2}}
\label{proofs}

% \subsection{Proof of Proposition~\ref{weak_limit_1}}

% We can decompose $S_{B,n}$ as
% $$
% S_{B,n} = \left| \sqrt{n} (\hat \xi_{B,n} - \xi_{B,\th_0}) - \sqrt{n} (\xi_{B,\hat \th_n} - \xi_{B,\th_0}) \right|.
% $$
% The first term in the absolute value converges in distribution to $\Lambda_B$. From the delta method, we have that $\sqrt{n} (\xi_{B,\hat \th_n} - \xi_{B,\th_0}) - \dot \xi_{B,\th_0}^\top \sqrt{n} (\hat \th_n - \th_0)$ converges to zero in probability. It follows that $S_{B,n}$ and $ | \sqrt{n} (\hat \xi_{B,n} - \xi_{B,\th_0}) - \dot \xi_{B,\th_0}^\top \sqrt{n} (\hat \th_n - \th_0) |$ have the same weak limit, which, from the continuous mapping theorem, is $| \Lambda_B - \dot \xi_{B,\th_0}^\top \bm \Theta |$. \qed

%\subsection{Proof of Proposition~\ref{weak_limit_2}}

Let $(W_{1,1},\dots,W_{1,d}),\dots,(W_{m,1},\dots,W_{m,d})$ be a random sample of size $m = \lfloor \gamma n \rfloor$ from $C_{\th_0}$ independent of the available data. Furthermore, let $\check A_m$ be one of the three corrected nonparametric estimators of the Pickands dependence function considered in Section~\ref{nonparametric} computed from the pseudo-observations obtained from $(W_{1,1},\dots,W_{1,d}),\dots,(W_{m,1},\dots,W_{m,d})$, and let $\check \xi_{B,m} = |B| \check A_m(\vec w_B)$ be the corresponding estimator of $\xi_{B,\th_0}$. Then, from the assumptions, we have that 
$$
 \left( \sqrt{m} (\check \xi_{B,m} - \xi_{B,\th_0}),  \frac{1}{\sqrt{m}} \sum_{i=1}^m \frac{\dot c_{\th_0}(W_{i,1},\dots,W_{i,d})}{c_{\th_0}(W_{i,1},\dots,W_{i,d})} \right)
$$
converges in distribution to $(\Lambda_B',\Wb')$, an independent copy of $(\Lambda_B,\Wb)$. It follows that
\begin{equation}
\label{sample_W}
 \left( \sqrt{n} (\check \xi_{B,m} - \xi_{B,\th_0}),  \frac{1}{\sqrt{n}} \sum_{i=1}^m \frac{\dot c_{\th_0}(W_{i,1},\dots,W_{i,d})}{c_{\th_0}(W_{i,1},\dots,W_{i,d})} \right)
\end{equation}
converges in distribution to $(\gamma^{-1/2} \Lambda_B', \gamma^{1/2} \Wb')$. Hence, by independence, we have that~\eqref{assumptions} and~\eqref{sample_W} converge jointly in distribution to $(\Lambda_B,\bm \Theta,\Wb,\gamma^{-1/2} \Lambda_B',\gamma^{1/2} \Wb')$. Consequently, the assumptions of the first part of Theorem~2 of \citet{GenRem08} are satisfied and we have that $\sqrt{n} ( \hat \xi_{B,n} - \xi_{B,\th_0}, \hat \xi_{B,\hat \th_n,m}  - \xi_{B,\th_0})$ converges in distribution to $(\Lambda_B, \gamma^{-1/2} \Lambda_B' + E(\gamma^{-1/2} \Lambda_B' \gamma^{1/2} \Wb'^\top ) \bm \Theta ) = (\Lambda_B, \gamma^{-1/2} \Lambda_B' + E(\Lambda_B \Wb^\top) \bm \Theta )$. 

Now, let us decompose $S_{B,n,m}$ as $S_{B,n,m} = \left|  \sqrt{n} (\hat \xi_{B,n} - \xi_{B,\th_0}) - \sqrt{n} (\hat \xi_{B,\hat \th_n,m}  - \xi_{B,\th_0}) \right|$. By the continuous mapping theorem, it follows that $S_{B,n,m}$ converges in distribution to $| \Lambda_B - \gamma^{-1/2}\Lambda_B' -  \Ex(\Lambda_B \Wb^\top) \bm \Theta |$. 

It thus remains to verify that $\Ex(\Lambda_B \Wb) = \dot \xi_{B,\th_0}$. We shall only consider the case $\hat \xi_{B,n} = \Ep$, the case $\hat \xi_{B,n} = \Ecfg$ being similar. Since Conditions 2.1 and 4.1 of \citet{Seg12} are assumed to hold, from Theorem~1 of \citet{GudSeg12}, we have that
$$
\Lambda_B  = - |B| A_{\th_0}^2(\vec w_B) \int_0^1 \Cb(\vec u^{\vec w_B}) \frac{\dd u}{u}, 
$$ 
where $\Cb$ is the weak limit of the empirical copula process \citep[see e.g.][]{Seg12}, and $\vec u^{\vec w_B} = (u^{w_{B,1}},\dots,u^{w_{B,d}})$. Then, %using Fubini's theorem,
$$
\Ex(\Lambda_B \Wb) = - |B| A_{\th_0}^2(\vec w_B) \int_0^1 \Ex \{ \Cb(\vec u^{\vec w_B}) \Wb \} \frac{\dd u}{u}.
$$
Now, from \citet[page 1108]{GenRem08}, we have that $\Ex \{ \Cb(\vec u) \Wb \} = \dot C_{\th_0}(\vec u)$ for all $\vec u \in [0,1]^d$. It follows that 
$$
\Ex(\Lambda_B \Wb) = - |B| A_{\th_0}^2(\vec w_B) \int_0^1 \dot C_{\th_0}(\vec u^{\vec w_B}) \frac{\dd u}{u} = - |B| A_{\th_0}^2(\vec w_B) \frac{\partial}{\partial \th} \left\{ \int_0^1 C_\th(\vec u^{\vec w_B}) \frac{\dd u}{u} \right\} \Big|_{\th=\th_0}
$$
where the last equality is a consequence of the continuity of $\th \mapsto \dot C_\th$,~(A3) and Lebesgue's dominated convergence theorem. Finally, from Lemma~1 of \citet{GudSeg12} \citep[see also][Lemma 3.1]{GenSeg09}, we have that $\int_0^1 C_\th(\vec u^{\vec w_B}) u^{-1} \dd u = 1/A_\th(\vec w_B)$, from which we obtain that $\Ex(\Lambda_B \Wb) = |B| \dot A_{\th_0}(\bm w_B) = \dot \xi_{B,\th_0}$. \qed

\section{Reducing the computational cost of the parametric bootstrap}
\label{compcost}

The parametric bootstrap is clearly a computationally intensive statistical procedure. Besides the fact that random number generation and fitting of the hypothesized model are necessary at each iteration, its high cost may additionally come from the cost of the evaluation of the estimate of the quantity of interest under the null. A strategy for speeding-up the procedure then consists of precomputing a reasonably accurate approximation of the function mapping the parameter vector to the quantity of interest under the null.

To fix ideas, let us focus on the algorithm given in Section~\ref{onelevel}. From Step 3~(c), we see that, for every $k \in \{1,\dots,N\}$, once $\hat \th_n^{(k)}$ is computed by fitting the hypothesized model to the data generated in Step 3~(a), $\xi_{B,\hat \th_n^{(k)}}$ needs to be evaluated so that $S_{B,n}^{(k)}$ can be computed. The last step is not necessarily straightforward even if a closed-form expression for the map $\th \mapsto \xi_{B,\th}$ is available. A good example of the latter fact is when the Smith model is hypothesized as the evaluation of~\eqref{extremal_Smith} turns out to be very costly. In such a situation, the speed of the parametric bootstrap procedure can be increased by precomputing a reasonably accurate approximation of the map $\th \mapsto \xi_{B,\th}$. It is however important to note that, in the context of max-stable processes, this last step may only be of interest in the framework of a simulation study as the map to be precomputed depends on the location of the $d$ sites.

\begin{figure}[tbp]
  \centering
  \includegraphics[width=\textwidth]{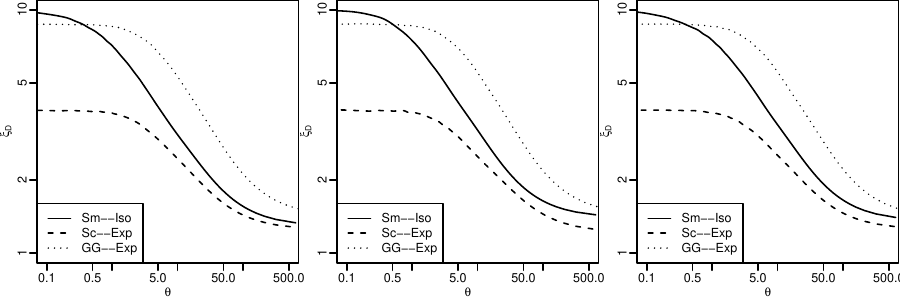}
  \caption{The left (resp.\ middle, right) plot represents the graph of the precomputed approximation of the mapping $\theta \mapsto \xi_{D,\theta}$  based on the CFG estimator for each of the three models in the case of the set of sites represented in the left (resp.\ middle, right) plot of Figure~\ref{fig:sites}.}
  \label{fig:ex10}
\end{figure}

A similar strategy can actually be used even if a closed-form expression for the map $\th \mapsto \xi_{B,\th}$ is unavailable. %and relies on the use of an estimator and results in an hydbrid between the one-level and the two-level parametric bootstrap if one is ready to neglect the additional randomness introduced by the second level of random number generation of the algorithm and treat that part as a numerical procedure for computing the estimate of the quantity of interest under the null. In the case of the algorithm given in Section~\ref{twolevel}, a comparison between Propositions~\ref{weak_limit_1} and~\ref{weak_limit_2} suggests that the influence of the second level of the parametric bootstrap can be made arbirarily small by taking $\gamma$ large. In practice, this means taking $m$ very large compared to $n$. 
Let us illustrate the proposed approach in the case of the simulations that produced Tables~\ref{tab:sm1}-\ref{tab:sm3}. For each of the three site configurations represented in Figure~\ref{fig:sites}, each set $B \subset D$, $|B| \in \{3,d\}$ and each of the three models Sm--Ani, Sc--Exp and GG--Exp parametrized by $\theta > 0$ as explained in Section~\ref{simulations}, a grid of $\theta$ values was created as $\theta = \arctan(\pi u/2)$ for $u \in \{0.001,0.002,\dots,0.999\}$. For each $\theta$ value on the grid, a sample of size $m=2500$ was generated under the model and the value of $\xi_{B,\theta}$ was estimated by $|B| \hat A_m(\vec w_B)$, where $\hat A_m$ is one of the three corrected nonparametric estimators of the Pickands dependence function defined in Section~\ref{nonparametric}. The relationship between the $\theta$ values and the corresponding $\xi_{B,\theta}$ values was approximated using penalized splines as implemented in the {\tt pspline} \textsf{R} package \citep{pspline} and stored for future use. An an example, the precomputed approximations of the mappings $\theta \mapsto \xi_{D,\theta}$ when the CFG estimator is used for $\hat A_m$ are represented in Figure~\ref{fig:ex10} for each of the three sets of sites represented in Figure~\ref{fig:sites}. 

Note that, although the precomputing step has some similarity with the second level of the algorithm of the two-level parametric bootstrap given in Section~\ref{twolevel}, the simulation procedure based on the precomputed approximations is indeed a one-level parametric bootstrap as the use of the latter does not bring in any additional variability. 

\begin{figure}[tbp]
  \centering
  \includegraphics[width=\textwidth]{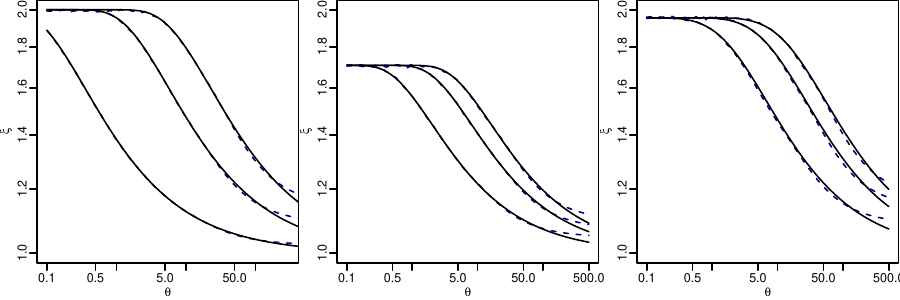}
  \caption{The left (resp.\ middle, right) plot compares the graphs of the mappings based on closed-form expressions (solid lines) with those of the corresponding precomputed approximations (dashed lines) based on penalized splines for the Sm--Iso (resp.\ Sc--Exp, GG--Exp) model. The top (resp.\ middle, bottom) pair of curves corresponds to fictitious sites at distance 1 (resp.\ 4, 8).}
  \label{fig:ex2}
\end{figure}

For $|B|=2$, the mappings $\theta \mapsto \xi_{B,\theta}$ were computed using the closed-form expressions available for all three models. The latter were also used to empirically validate the accuracy of the procedure producing the approximations of the precomputed mappings. As an illustration, Figure~\ref{fig:ex2} compares the graphs of the mappings based on closed-form expressions with those of the corresponding precomputed approximations based on penalized splines for the three models used in the simulations. As one can see, the approximations appear reasonably accurate except when $\theta$ is very large.

\bibliographystyle{plainnat}
\bibliography{biblio}

\end{document}